\documentclass[10pt, a4]{article}

\usepackage{graphicx, hyperref}
\usepackage{amsfonts, amsmath,amssymb, mathrsfs, wasysym}
\usepackage[english]{babel}
\usepackage[normalem]{ulem}
\usepackage{color,slashed}

\ifx\pdfoutput\undefined
\usepackage[dvips,bookmarks]{hyperref}	
\else
\usepackage{hyperref}	
\fi
\hypersetup{colorlinks,bookmarksopen,bookmarksnumbered,citecolor=rossos,
linkcolor=rossos,pdfstartview=FitH,urlcolor=rossos}
\newcommand{\Tr}{{\rm Tr}\,}

\def\circa#1{\,\raise.3ex\hbox{$#1$\kern-.75em\lower1ex\hbox{$\sim$}}\,}

\numberwithin{equation}{section} 
\usepackage[table]{xcolor}

\definecolor{grigino}{cmyk}{0,0,0,0.2}
\definecolor{mentuccia}{cmyk}{0.4,0,0.3,0.1}
\definecolor{arancino}{cmyk}{0,0.1,0.4,0}
\definecolor{menta}{cmyk}{0.7,0,0.5,0.3}
\definecolor{grigios}{cmyk}{0,0,0,0.5}
\definecolor{bianco}{cmyk}{0,0,0,0}
\definecolor{arancio}{cmyk}{0,0.2,0.6,0}
\definecolor{grigio}{cmyk}{0,0,0,0.1}
\definecolor{rosa}{cmyk}{0,0.1,0.1,0.02}
\definecolor{rosino}{cmyk}{0,0.05,0.05,0.02}
\definecolor{rosas}{cmyk}{0,0.3,0.25,0.05}
\definecolor{celeste}{cmyk}{0.1,0,0,0.02}
\definecolor{giallino}{cmyk}{0,0,0.4,0.02}
\definecolor{rosso}{cmyk}{0,1,1,0.4}
\definecolor{rossos}{cmyk}{0,1,1,0.55}
\definecolor{rossoc}{cmyk}{0,1,1,0.2}
\definecolor{blu}{cmyk}{1,1,0,0.3}
\definecolor{blus}{cmyk}{1,1,0,0.5}
\definecolor{bluc}{cmyk}{1,1,0,0.1}
\definecolor{blucc}{cmyk}{0.7,0.5,0,0}
\definecolor{viola0}{cmyk}{0,0.4,0,0.04}
\definecolor{viola}{cmyk}{0,0.5,0,0.05}
\definecolor{viola2}{cmyk}{0,1,0.2,0.6}
\definecolor{verde}{cmyk}{0.92,0,0.59,0.25}
\definecolor{verdec}{cmyk}{0.92,0,0.59,0.15}
\definecolor{verdecc}{cmyk}{0.42,0,0.8,0.05}
\definecolor{verdes}{cmyk}{0.92,0,0.59,0.4}
\definecolor{verdino}{cmyk}{0.12,0,0.3,0.02}
\definecolor{giallo}{cmyk}{0,0,1,0}
\definecolor{gialloverde}{cmyk}{0.44,0,0.74,0}

\newcommand{\be}{\begin{equation}}
\newcommand{\ee}{\end{equation}}
\newcommand{\bea}{\begin{eqnarray}}
\newcommand{\eea}{\end{eqnarray}}

\newcommand{\bu}{{\bf 1}}

\newcommand{\newc}{\newcommand}

\newc{\gsim}{\lower.7ex\hbox{$\;\stackrel{\textstyle>}{\sim}\;$}}
\newc{\lsim}{\lower.7ex\hbox{$\;\stackrel{\textstyle<}{\sim}\;$}}

\numberwithin{equation}{section}

\def\fun#1#2{\lower3.6pt\vbox{\baselineskip0pt\lineskip.9pt
  \ialign{$\mathsurround=0pt#1\hfil##\hfil$\crcr#2\crcr\sim\crcr}}}
\def\simgt{\mathrel{\lower0.6ex\hbox{$\buildrel {\textstyle >}
 \over {\scriptstyle \sim}$}}}
\def\simlt{\mathrel{\lower0.6ex\hbox{$\buildrel {\textstyle <}
 \over {\scriptstyle \sim}$}}}
\def\mK{{\mathcal K}}
\def\bea{\begin{eqnarray}}
\def\eea{\end{eqnarray}}
\def\be{\begin{equation}}
\def\ee{\end{equation}}
\def\tr{{\rm tr}\,}

\input epsf

\def\be{\begin{equation}}
\def\ee{\end{equation}}
\def\ba{\begin{eqnarray}}
\def\ea{\end{eqnarray}}

\def\im{\eta^{-1}}

\begin{document}

\title{
\vspace{-2.0cm}
\vspace{2.0cm}
{\bf \huge  The Self-Accelerating Universe with Vectors in Massive Gravity
}
 \\[8mm]
}

\author{
Kazuya Koyama,\, Gustavo Niz,\, Gianmassimo Tasinato
\\[8mm]
\normalsize\it
Institute of Cosmology \& Gravitation, University of Portsmouth,\\
\normalsize\it Dennis Sciama Building, Portsmouth, PO1 3FX, United Kingdom
}

\date{}
\maketitle

\thispagestyle{empty}

\begin{abstract}
\noindent
We explore the possibility of realising self-accelerated expansion of
the Universe taking into account the vector components of a massive graviton.
The effective action in the decoupling limit  
contains an infinite number of terms, once the vector degrees of freedom are included. These 
 can be re-summed in physically interesting situations, which result in non-polynomial
 couplings between the scalar and vector modes. We show there are self-accelerating
background solutions for this effective action, with the possibility of having a non-trivial profile for
 the vector fields. We then study fluctuations around these solutions and show 
 that there is always a ghost, if a background vector field is present. When the background vector field is switched off, the ghost can be avoided, at the price of entering into a 
  strong coupling regime, in which the vector fluctuations have vanishing kinetic terms. Finally we show that the inclusion of a bare cosmological constant does not change the previous conclusions and it does not lead to a ghost mode
in the absence of a background vector field. 

  
 \end{abstract}


\def\eq#1{eq.~(\ref{#1})}

\bigskip

\section{Introduction}
\label{Introduction}

Understanding the nature of dark energy represents one of the most interesting open questions in cosmology. Modifications of Einstein's theory of gravity at large scales could explain the present day acceleration, without invoking the aid of a cosmological constant or a exotic matter content. Therefore, self-interactions in the gravitational sector may be sufficient to realise cosmological acceleration with no need of an additional energy momentum tensor; this phenomenon is dubbed self-acceleration.

Massive gravity is an example of these modified gravity models, in which the self-acceleration is realized thanks to the specific dynamics of gravitational degrees of freedom. Already in the Fierz-Pauli version of massive gravity
\cite{Fierz:1939ix}, explicit solutions were found describing a de Sitter spacetime \cite{Salam:1976as}, with the de Sitter radius inversely proportional to the graviton's mass. Recently, in the context of a novel non-linear extension of Fierz-Pauli \cite{deRham:2010ik,claudia}, self-accelerating solutions have been discovered \cite{us, us2, Nieuwenhuizen:2011sq, deRham:2010tw, D'Amico:2011jj, Gumrukcuoglu:2011ew} [See also \cite{ArkaniHamed:2002sp, Hinterbichler:2011tt, david, Hassan:2011hr, Chamseddine:2010ub} for a non-exhaustive list of work in this theory and related models].
In a healthy theory of massive gravity, the graviton contains five degrees of freedom: two tensors, two vectors, and one scalar. The dynamics of the scalar mode has been studied in detail in \cite{deRham:2010tw}, showing that a non-trivial configuration for this field leads to self-acceleration. The analysis was carried out in the so-called decoupling limit, a regime of linearised dynamics in the tensor modes, and in which the scalar-scalar, as well as scalar-tensor interactions, are described by a finite number of terms, which can be transformed into the so-called Galileon combinations \cite{Nicolis:2008in}. Scalar fluctuations around these self-accelerating configurations are free of ghosts, provided that parameters characterizing the theory are chosen appropriately \cite{deRham:2010tw}. On the other hand, around this self-accelerating configuration, the vector degrees of freedom enters into a regime of strong coupling, as the coefficient in front of the kinetic term vanishes \cite{deRham:2010tw}.

However, in all the literature of self-accelerating solutions in massive gravity, little attention has been put into the vector mode. One may ask if the theory admits other classes of self-accelerating configurations, which include vectors, and if they exist what is their perturbative stability.
In this paper we answer these questions, presenting new solutions in which the {\it vector} degrees of freedom play a crucial role, supporting, together with the scalar mode, the self-acceleration. We focus  on the theory in the decoupling limit, and develop tools to obtain an effective action describing the vector modes in this limit. This effective action turns out to have an infinite number of terms for the vector-scalar interactions, in contrast to the finite number for the non-vectorial sector. This infinite series leads to non-polynomial interactions between the scalar and the vector. By concentrating our attention in spherically symmetric ans\"atze, we are able to resum the series, and find general self-accelerating solutions in the effective theory  which describe  de Sitter space  with non-trivial profiles for the scalar and vectors. These general solutions, as far as we know, represent 
the first example of self-accelerating configurations with a vector field. We then proceed to study the dynamics of fluctuations around these general solutions. We find that a non-trivial profile of the vector field is able to alleviate the strong coupling behaviour of vector fluctuations, by providing positive definite kinetic terms to them. However we find that one of the scalar and vector perturbations always becomes a ghost, so that the only ghost-free self-accelerating solutions are those without the background vector field. Furthermore, a bare cosmological constant can be included in these self-accelerating solutions and the theory of fluctuations about them remains ghost-free if the background vector mode is not present. However, in the case of a non-trivial profile for the background vector field, the ghost mode cannot be avoided. 

The paper is organized as follows. In Section \ref{sec-massgrav}, we review the construction of the non-linear Lagrangian for massive gravity. In Section \ref{sec-selfacc}, we focus on the simplest version of this theory, presenting new self-accelerating solutions with a vector field included. In Section \ref{Lagsection}, we determine the Lagrangian for massive gravity in the decoupling limit including vector degrees of freedom. In Section \ref{secstabsim}, we study the dynamics of linear fluctuations around the self-accelerating configurations, showing that a ghost is always present in the spectrum if there is background vector. In Section \ref{sec-a3a4}, we extend our analysis to the full theory including two additional parameters and determine the most general self-accelerating solutions in the decoupling limit with vector fields. The analysis of linear fluctuations around these configurations show that there is no parameter space where we can remove the ghost if a background vector field is switched on. We present our conclusions  in Section 7. Three Appendixes complete the
paper with technical details. 

\section{Massive Gravity Lagrangian}\label{sec-massgrav}

We consider the following Lagrangian for massive gravity, a non-linear
extension of Fierz-Pauli theory proposed in \cite{claudia}
\be\label{genlag}
{\cal L} = \frac{M_{Pl}^2}{2}\,\sqrt{-g}\left( R -2\Lambda - {\cal U}\right).
\ee
The potential depends on a dimension-full  parameter $m$, which sets the graviton 
mass scale, and on two dimensionless parameters $\alpha_3$ and $\alpha_4$. 
It has the following form:
\be
{\cal U}= -m^2\left[{\cal U}_2+\alpha_3\, {\cal U}_3+\alpha_4\, {\cal U}_4\right],
\label{potentialU}
\ee
with 
\bea
{\cal U}_2&=&(\tr\mK)^2-\tr (\mK^2),\nonumber \\
{\cal U}_3&=&(\tr \mK)^3 - 3 (\tr \mK)(\tr \mK^2) + 2 \tr \mK^3,\nonumber \\
{\cal U}_4 &=& (\tr \mK)^4 - 6 (\tr \mK)^2 (\tr \mK^2)
+ 8 (\tr \mK)(\tr \mK^3) + 3 (\tr \mK^2)^2 - 6 \tr \mK^4 .\nonumber
\eea
The tensor ${\cal K}_{\mu}^{\ \nu}$ is defined as \cite{claudia}
\bea
{\mathcal K}_{\mu}^{\ \nu} &\equiv &\delta_{\mu}^{\ \nu}-\left(\sqrt{g^{-1} \left[g-h\right]}\right)_{\mu}^{\ \nu}\,\,,
\eea
where the square root of a tensor is defined as $\sqrt{{\cal M}}_{\mu}^{\ \alpha}\sqrt{\cal M}_{\alpha}^{\ \nu}={\cal M}_{\mu}^{\ \nu}$,
for any tensor ${\cal M}_\mu^{\,\,\nu}$. The metric $h_{\mu\nu}$ is the displacement from a fiducial flat metric $\eta_{\mu\nu}$, namely $h_{\mu \nu} \equiv \eta_{\mu\nu} -g_{\mu\nu}$. The action (\ref{genlag}) breaks explicitly reparametrisation invariance, but it can be restored by means of the St\"uckelberg trick \cite{ArkaniHamed:2002sp}. The new definition for the tensor ${\cal K}_{\mu}^{\ \nu}$ is given by 
\bea\label{covK}
{\mathcal K}_{\mu}^{\ \nu} &\equiv &\delta_{\mu}^{\ \nu}-\left(\sqrt{g^{-1} \left[g-H\right]}\right)_{\mu}^{\ \nu}\,\,,
\eea
where now $H_{\mu \nu}$ corresponds to the covariantisation of the metric perturbations, defined as 
\bea
H_{\mu \nu}\equiv h_{\mu\nu}+\eta_{\beta \nu}\partial_\mu \pi^\beta+\eta_{\alpha \mu}\partial_\nu \pi^\alpha- \eta_{\alpha \beta} \partial_\mu \pi^\alpha \partial_\nu \pi^\beta.
\eea
Therefore, a change of coordinates $x^{\mu} \to x^{\mu} + \xi^{\mu}$ should be accompanied by the following transformation of the St\"uckelberg field $\pi^\mu$,
\be\label{pi_trans}
\pi^{\mu} \to \pi^{\mu} + \xi^{\mu},
\ee
in order to recover full diffeomorphism invariance.
The field $\pi^\mu$ can be decomposed into a divergence-less vector $A_\mu$, and a scalar $\pi$ in the following way
\be\label{decomp}
\pi^\mu=\eta^{\mu\nu}(A_\nu+\partial_\nu \pi).
\ee
Consequently, in terms of the vector and scalar, the tensor $H_{\mu\nu}$ can be expressed as
\be\label{Hdef}
H_{\mu\nu}=h_{\mu\nu}+2\Pi_{\mu\nu}-\Pi^2_{\mu\nu}+\partial_\mu A_\nu+\partial_\nu A_\mu-\partial_\mu A^\alpha \partial_\nu A^\beta\eta_{\alpha\beta}-\Pi_{\mu\alpha}\partial_\nu A^\alpha-\partial_\mu A^\alpha\Pi_{\nu\alpha},
\ee
where $\Pi_{\mu\nu}^2=\Pi_{\mu\alpha}\Pi^{\alpha}_\nu$ and $\Pi_{\mu\nu}=\partial_\mu\partial_\nu\pi$. The indices of $\pi^\mu$ are raised/lowered by the flat fiducial metric $\eta_{\mu\nu}$.

\smallskip
The St\"uckelberg scalar  $\pi$ and vector $A^\mu$ play the roles of scalar and vector components of the massive graviton: the Lagrangian obtained by applying the St\"uckelberg trick contains the same number of degrees of freedom as the original Lagrangian (\ref{genlag}). By plugging the expression for the tensor ${\cal K}_\mu^{\,\,\nu}$, written  in terms of  $h$, $A$ and $\pi$, inside (\ref{genlag}), one obtains a Lagrangian for the massive gravity theory expressed in terms of tensor, scalar, and vector degrees of freedom.
In order to canonically normalize the degrees of freedom\footnote{For the tensor $h_{\mu\nu}$ and vector $A_\mu$, the kinetic terms set the canonical normalization, whereas for the scalar field $\pi$ the kinetic terms are total derivatives. Given that one would like to keep some terms involving the scalar field in the decoupling limit (\ref{dec_lim}), we use normalization (\ref{cannorm}). This preserves non-vanishing scalar-tensor couplings in that limit, hence providing kinetic terms for the scalar after a diagonalisation. Similarly, the cosmological constant $\Lambda$ should scale as $h_{\mu\nu}$ in order to have consistent solutions in the decoupling theory.}, 
we need to rescale the fields as (see for example \cite{Hinterbichler:2011tt})
\be\label{cannorm}
h_{\mu\nu}\,\to\,M_{Pl}\, h_{\mu\nu}
\hskip 0.5cm , \hskip 0.5cm A_{\mu}\,\to\,m M_{Pl}\, A_\mu
\hskip 0.5cm , \hskip 0.5cm \pi\,\to\,m^2 M_{Pl}\, \pi,
\hskip 0.5cm , \hskip 0.5cm \Lambda\,\to\,M_{Pl}\, \Lambda.
\ee
The transformations associated with the diffeomorphism invariance Eq.~(\ref{pi_trans}) can also be rewritten in terms of these fields as
\bea
\delta  h_{\mu \nu}&=&\partial_\mu \xi_\nu + \partial_\nu \xi_\mu+\frac{2}{M_{Pl}}\,{\mathcal L}_{\xi}\,
 h_{\mu\nu},
\nonumber\\
\delta  A_{\mu}&=&\partial_\mu \lambda-m\, \xi_{\mu}+\frac{ 2\, \xi^\nu\, \partial_\nu\, A_{\mu}}{M_{Pl}},
\nonumber\\
\delta  \pi&=&-m \,\lambda\label{diffeocanon},
\eea
for an arbitrary divergence-less vector $\xi_\mu$ and a scalar  $\lambda$. This Lagrangian   contains various non-linear interactions characterized by derivative  couplings, which in turn play a crucial role for allowing an implementation of the Vainshtein  mechanism \cite{us, us2,vanDam:1970vg, david}, and for generating self-accelerating configurations leading to the de Sitter expansion \cite{deRham:2010ik, us, us2, deRham:2010tw, D'Amico:2011jj, Gumrukcuoglu:2011ew}. 

 \smallskip

The aim of this paper is to discuss  new  self-accelerating solutions for massive gravity, exploiting the features
of the vector sector of the theory. In order to do so,
it is convenient to focus in
the decoupling limit \cite{ArkaniHamed:2002sp} 
\be\label{dec_lim}
m\,\to\,0\hskip 0.5cm , \hskip 0.5cm M_{Pl}\,\to\,\infty\hskip 0.5cm , \hskip 0.5cm \Lambda_3\,\equiv\,
m^2 M_{Pl}
\,=\,{\rm fixed}.
\ee
This limit is particularly appropriate for our objectives, because the tensor piece in the Lagrangian simplifies considerably. On one hand, the tensor self-interactions are described by the quadratic expansion of Einstein-Hilbert action, while interactions with other degrees of freedom are linear in the tensor mode. On the other hand, the Lagrangian
describing vector and scalar degrees of freedom maintains its non-linear structure, admitting de Sitter solutions with no need of an additional energy momentum tensor \cite{deRham:2010ik,us,us2}. In these solutions, the de Sitter radius is proportional to the inverse of the strong coupling scale $\Lambda_3$.

The Lagrangian in the decoupling limit of Eq.~(\ref{dec_lim}) has a reduced symmetry with respect to (\ref{diffeocanon}), namely
\bea
 \delta  h_{\mu \nu}&=&\partial_\mu \xi_\nu + \partial_\nu \xi_\mu,
\nonumber\\
\delta  A_{\mu}&=&\partial_\mu \lambda
\nonumber,\\
\delta  \pi&=&0, 
\label{diffeodeco}
\eea
hence the symmetry is reduced to the linearised diffeomorphism invariance times an independent $U(1)$ symmetry for the divergence-less vector.
The Lagrangian in the decoupling limit, when expanded from a flat fiducial metric,
has been shown to be free from the so-called Boulware-Deser (BD) ghost \cite{Boulware:1973my, deRham:2010ik}, in the sense that it contains only five degrees of freedom; the sixth mode, potentially leading to a ghost, is absent \footnote{When one considers the fiducial flat metric as the background for studying linear fluctuations, there is no way of exciting a possible sixth degree of freedom using the vector-scalar coupling, since the action in the decoupling limit is quadratic in the vector field. However, as we show in this paper, vector fluctuations play an important role around different backgrounds.}. However, around the self-accelerating de Sitter configurations, one of the remaining physical five degrees of freedom may become a ghost. As we will learn in this paper, there are conditions on the parameters of the theory  in order to have a set-up that is ghost free around the self-accelerating configurations.

There are strong indications that the theory is ghost free (in the sense that the
Boulware-Deser sixth mode is absent) also away from the decoupling limit \cite{Hassan:2011hr}, although there is still debate on this issue \cite{Creminelli:2005qk,Alberte:2010it,Gruzinov:2011sq, deRham:2011rn}. The analysis of this subject 
is not the main aim of this work, so we will not discuss it here. 

In the next section, we present a method for determining new self-accelerating configurations, in the decoupling limit of this massive gravity theory, but including the dynamics of vector degrees of freedom. We determine them in two ways: firstly, we show that they can be obtained by taking the decoupling limit of known exact solutions in the full theory. Secondly, by imposing the appropriate symmetries, we derive an effective action for the relevant degrees of freedom (including vectors) whose general solution provides the same self-accelerating configurations.

\section{Self-accelerating solutions with vectors ($\alpha_3=\alpha_4=0$ case)}\label{sec-selfacc}

In order to study new self-accelerating solutions in the presence of vector degrees of freedom, we proceed as follows: we start with the most general static spherically symmetric solutions for the original Lagrangian (\ref{genlag}). As this Lagrangian stands, it describes only the dynamics of the metric $h_{\mu\nu}$, and as we previously showed \cite{us,us2}, there is a branch of solutions exhibiting the static patch of the de Sitter spacetime.

By reintroducing the diffeomorphism invariance via the St\"uckelberg trick (\ref{covK}), we apply a gauge transformation that allows to recast the fields in a form that is suitable for taking the decoupling limit (\ref{dec_lim}). The details on how this procedure is performed are in Appendix B. After this limit is taken, we get a self-accelerating configuration where the de Sitter invariance is fully manifest. Moreover, we find that these solutions include a non-trivial profile for the vector field. As far as we are aware, this is the first example of self-accelerating configurations which includes vector modes.

We then proceed to show that the same class of self-accelerating configurations can be alternatively obtained as solutions for the Lagrangian in the decoupling limit, imposing 
an appropriate Ansatz for the fields involved that respects the symmetries of the de Sitter spacetime.

\subsection{Self-acceleration from spherically symmetric solutions}

 \smallskip

The problem of  classifying the most general spherically
symmetric static solutions for the  original, non-covariant, Lagrangian (\ref{genlag}), describing the dynamics of tensor degrees of freedom, has been already discussed in the literature \cite{us, us2, Gruzinov:2011mm}. If de Sitter solutions exist, their static patches (if any) have to be contained within our classification. Here we briefly review our findings.  We start with the most general Ansatz for static spherically symmetric configurations,
\be\label{genmetr}
d s^2\,=\,-C(r) \,d t^2+A(r)\, d r^2 +2 D(r)\, dt dr+B(r) d \Omega^2,
\ee
where $d \Omega^2 = d \theta^2 + \sin^2 \theta d \phi^2$.
The non-dynamical flat metric is written in terms of spherical coordinates as
\be\label{flspher}
ds_{flat}^2 = -dt^2 + dr^2 + r^2 d \Omega^2\,.
\ee
Notice that in general relativity, one can set $B(r)=r^2$ by a coordinate transformation,
but this is not possible here, since we do not have diffeomorphism invariance.
In order to simplify our analysis, it is convenient to define the combination
$\Delta(r)\,=\,A(r) C(r)+D^2(r)$. We plug the previous metric into the Einstein equations
\be\label{einstein}
G_{\mu \nu}=T^{{\cal U}}_{\mu \nu},
\ee
where the energy momentum tensor associated
with the potential ${\cal U}$ of Eq. (\ref{potentialU}) is defined as $T^{{\cal U}}_{\mu \nu}\,=\frac{m^2}{\sqrt{-g}}\,\frac{ \delta \sqrt{-g} {\cal U}}{\delta g^{\mu \nu}}$. The Einstein tensor $G_{\mu\nu}$ satisfies the identity $D(r)\, G_{tt}+C(r)\,G_{tr}\,=\,0$, which implies
the algebraic constraint
 \bea
 0&=& D(r)\, T^{{\cal U}}_{tt}+C(r)\,T^{{\cal U}}_{tr} \nonumber\\
 &=& m^2\,\frac{
 D(r)\,
 \left(2 r -3 \sqrt{B(r)}\right)
 \,\sqrt{\Delta(r)}
 }{\sqrt{B(r)}\,\left( A(r)+C(r) +2\sqrt{\Delta(r)}\right)^{1/2}}\label{eqbranches}.
 \eea
The previous condition can be satisfied in two ways, which lead to {\it two} different branches of solutions \cite{us, Gruzinov:2011mm} (see also \cite{Damour:2002gp, pilo}).
We can either set $D(r)=0$, and focus on diagonal metrics, or alternatively set $B(r)\,=\,4\,r^2/9$. The fact that there are two branches of solutions indicates that, unlike in general relativity where the Birkhoff theorem holds,
there is no uniqueness theorem for spherically symmetric configurations in this theory. The diagonal branch gives solutions which are asymptotically flat \cite{us, us2}\footnote{The decoupling limit Lagrangian of this branch has only one asymptotic solution, which indeed decays to the flat space at a large distance from a source for the case $\alpha_3=\alpha_4=0$ \cite{us}. However, in the most general case, where $\alpha_3$ and $\alpha_4$ do not vanish, there are more asymptotic solutions than the flat space one \cite{us2}.}. Since we are interested in de Sitter configurations whose curvature invariants do not decay at infinity, we will not discuss further the diagonal branch of solutions. On the contrary, the second branch provides asymptotically de Sitter configurations.
To show this explicitly, we impose $B(r) =4\,r^2/9$.  Due to identity
$C(r) T^{{\cal U}}_{rr} + A(r) T^{{\cal U}}_{tt}=0$, one gets a further condition
\be
\Delta(r)  \,=\,A(r) C(r)+D^2(r) \,\equiv\,\Delta_0\,=\,\mathrm{const}.
\ee
The remaining Einstein equations provide the following {\it unique} solution
(see Ref.~\cite{us2} for detailed derivations)
\bea\label{gmcoeff}
A(r) &=& \frac{9 \Delta_0}{4} (p(r) + \alpha +1),\;\;\;\;\;
B(r) =\frac49 r^2,\\
C(r) &=&\frac{9 \Delta_0}{4}(1 - p(r)), \;\;
D(r) = \frac{9}{4} \Delta_0 \sqrt{p(r)(p(r)+\alpha)}, \nonumber
\eea
where
\be
p(r)=\frac{c}{r} + \frac{m^2 r^2}{9}, \;\;
\alpha =\frac{16}{81\,\Delta_0}-1,
\ee
with arbitrary constants $c$ and $\Delta_0$.
Notice that this configuration depends on {\it two} integration constants.
A sufficient condition to ensure that $D(r)$ is real, is to choose $c \geq 0$ and $0<\sqrt{\Delta_0} \leq 4/9$. The integration constant $c$ corresponds to the Schwarzschild mass, but since we are interested in pure de Sitter metrics, we set $c=0$ in what follows.
The general case ($\alpha_3,\, \alpha_4\neq0$) can be analysed in the same way, but its
discussion is postponed to Section \ref{sec-a3a4}. These solutions correspond to 
a maximally symmetric de Sitter space, since the Ricci tensor is proportional to the metric. If we perform coordinate transformations and go to the flat Friedman-Robertson-Walker slicing of the de Sitter spacetime, there appears a coordinate singularity at the de Sitter
horizon. The physical nature of these singularities is an interesting open issue, however since we focus in the decoupling limit, this singularity does not show up in our analysis. In fact, it was shown that there are no closed or flat FRW solutions in this theory \cite{D'Amico:2011jj}. However, a self-accelerating open FRW solution was recently found \cite{Gumrukcuoglu:2011ew}, and in the decoupling limit, this last solution is indeed included in the solutions that we obtain from exact spherically symmetric solutions.

\bigskip

As explained in the previous sections, we can restore the diffeomorphism invariance
via the St\"uckelberg trick, and apply a gauge transformation that recasts the metric 
of Eq.~(\ref{genmetr}) (with  $c=0$) in a form that makes the de Sitter nature of this space-time manifest. We developed this procedure in \cite{us, us2}, where we wrote the de Sitter spacetime in a explicitly time-dependent frame, while in what follows we will express it in conformally flat coordinates\footnote{Note that these coordinate transformations can unnecessarily introduce singularities in $\pi^{\mu}$. Again in the decoupling limit these singularities do not show up.}. Details on how to obtain the decoupling limit of solution (\ref{gmcoeff}) can be found in Appendix \ref{AppB}, but here we only describe the final result:
\bea
d s^2 &=&
 \left[
 1-\frac{\Lambda_3}{8\,M_{Pl}^2}(r^2-t^2) \right]\,(-d t^2 +d r^2 + r^2 d \Omega^2)+ {\cal O}(m^3)\,,\nonumber\\
 {A}_\mu&=&
\left(-\frac{\Lambda_3 r^2}{24}\sqrt{\frac{16}{\Delta_0}-81}+{\cal O}(m^2) ,0,0,0\right)\,,\nonumber \\
 \pi&=& -\frac{\Lambda_3}{4}(r^2-t^2) +\frac{3\,\Lambda_3}{4}\left(\frac{4}{9\,\sqrt{\Delta_0}}-1\right)t^2
+{\cal O}(m^2) \,.\label{vectorsol}
\eea
where we used $M_{Pl}=\Lambda_3/m^2$ and only wrote the term at leading order in $m$,
since we intend to focus in the decoupling limit (\ref{dec_lim}).
Indeed in this particular limit, we take $m \to 0$, so only the
terms we explicitly wrote survive. The metric then assumes a manifestly de Sitter form in a conformally flat frame. This configuration provides a generalization of the solution
analysed in \cite{deRham:2010tw} where only the scalar field is switched on. Actually, the choice $\Delta_0\,=\,16/81$ switches off the vector and the solution reduces precisely to that of \cite{deRham:2010tw}. However, in general also the vector degree of freedom is present.
Moreover, it is important to stress that even though we derived these solutions (\ref{anddec}) from the specific non-linear solution (\ref{gmcoeff}), the decoupling limit solutions are more general in the sense that many different non-linear solutions can be reduced to the same decoupling solutions (\ref{vectorsol}). For example and as mentioned before, the open FRW solution obtained in Ref.~\cite{Gumrukcuoglu:2011ew} reduces to (\ref{vectorsol}) with $\Delta_0\,=\,16/81$ in the decoupling limit. We will study various full non-linear solutions describing the self-accelerating universe in a future publication \cite{new2}.


\subsection{Solving the field equations in the decoupling limit}

It is also possible to reproduce the same configurations by solving the equations of motion for the Lagrangian in the decoupling limit by imposing the appropriate symmetries. In order to maintain spherical symmetry, we adopt the following Ansatz for the canonically normalized fields.
\bea\label{anddec}
g_{\mu\nu}&=&\left(1+\frac{f(t,r)}{M_{Pl}}\right)\,\eta_{\mu\nu},\nonumber \\
A_\mu&=&(A_t(r),0,0,0),\nonumber \\
\pi&=& \pi_t(t)+ \pi_r(r).
\eea
This Ansatz is able to generate the de Sitter metric expressed in conformally flat coordinates. Although it is not the most general Ansatz compatible with the required
symmetries, it is sufficient for reproducing the solution
(\ref{vectorsol}).

After inserting the Ansatz (\ref{anddec}) in the action
(\ref{genlag}), we find the following Lagrangian
in the decoupling limit for the fields $f$, $A_t$ and $\pi$ (with the fiducial flat metric expressed as (\ref{flspher})),
\be\label{dec_lag}
{\cal L}_{dec}=\frac{r^2}{2}\left[ \frac{F_1( \pi )}{m^2}+{\cal L}_R^{(2)}({f})+F_2( \pi){f}(t,r)+F_3(\pi)(\partial_r {A}_t)^2\right]\,,
\ee
where
\bea
F_1( \pi)&=&\left(2 \pi'' + \frac{\pi'}{r}\right) \frac{\pi'}{r} - \left(\pi''
 + \frac{2\,\pi'}{r}\right) \ddot{\pi} 
 \nonumber ,\\
F_2( \pi)&=& 2\left(3 +  \frac{\pi'}{r}\right)  \frac{\pi'}{r} + \left(3 +  \frac{4\,\pi'}{r}\right) \pi'' - \left(3 + 2 \pi''
 + 4  \frac{\pi'}{r}\right) \ddot{\pi}
\nonumber ,\\
F_3(\pi)&=&\frac{r+2 \pi'}{r \left(2+ \ddot{\pi}-\pi'' \right)}\, .
\eea
The dot and the prime correspond to derivatives along $t$ and $r$ respectively. Moreover,
we set $\Lambda_3 =1$ for convenience.
The $r^2F_1/m^2$ term is a total derivative, thus it does not contribute to the equations of motion. The second term, ${\cal L}_R^{(2)}$, is the Einstein-Hilbert action truncated to quadratic order in $ f$. The direct coupling between $ f$ and $ \pi$ ({\it i.e.} $F_2(\pi)\, f$) corresponds to the contribution found in \cite{deRham:2010ik}, and denoted there
by $h^{\mu\nu}(X^{(1)}_{\mu\nu}+X^{(2)}_{\mu\nu})$, where  $X^{(n)}_{\mu\nu}$ has $n$ powers of $\partial_\mu\partial_\nu\pi$. This coupling leads to the second order differential equation for the field $ \pi$.

Regarding the vector sector, we obtain a non-polynomial coupling between the vector and scalar degrees of freedom controlled by the function $F_3$. Derivatives of the scalar 
field appear in the denominator of this expression {\it even in the decoupling limit}. 
We will see that this is a result of a resummation of an infinite number of terms. 
We will discuss how to obtain these couplings in general in Section \ref{Lagsection}. 
The equation of motion for the scalar mode, $\pi$, is given by
$$\partial_t^2\left(\frac{\delta {\cal L}}{\delta \ddot\pi}\right)+\partial_r^2\left(\frac{\delta {\cal L}}{\delta \pi''}\right)-\partial_r\left(\frac{\delta {\cal L}}{\delta \pi'}\right)\,=\,0\,.$$
The first term in the previous expression could lead to higher order derivatives in time if
$\delta {\cal L}/\delta \ddot \pi$ depends on $\ddot \pi$. The $F_2$ coupling in the Lagrangian will not generate such higher derivatives, in agreement with the findings of \cite{deRham:2010ik}. In contrast, the vector-scalar coupling $F_3$ might give rise to such higher derivative terms, which might be thought as the propagation of the sixth-degree of freedom. However, as we will see later, in the equations of motion describing fluctuations around our self-accelerating solutions all these higher derivative terms cancel exactly, ensuring that the Boulware-Deser mode is absent.

\smallskip

The equations of motion associated with Lagrangian (\ref{dec_lag}) have the following asymptotically non-decaying solution 
\bea\label{solwq0}
 f&=&-\frac{\Lambda_3}{8}\,\left(r^2-t^2\right)\,,\nonumber\\
 A_t&=&-\frac{\Lambda_3\,Q_0}{2}\,r^2\,,\nonumber\\
 \pi&=&-\frac{\Lambda_3}{4}\,\left(r^2-t^2\right)+\frac{3\,\Lambda_3}{4}\,\left(\sqrt{1+\frac{16 Q_0^2}{9}}-1\right)\,t^2
\eea
where $Q_0\geq 0$ is a free integration constant.
It is straightforward to check that, choosing $Q_0\,=\frac{1}{12} \sqrt{\frac{16}{\Delta_0}-81}$, the previous solution
indeed reduces to (\ref{vectorsol}) when $m\to 0$. This proves that the configuration  (\ref{vectorsol}) is a solution of the field equations in the decoupling limit.

\section{The Lagrangian with vector degrees of freedom}\label{Lagsection}

We are now interested in deriving a more general expression for the Lagrangian in the decoupling limit including vector degrees of freedom. As we will learn, this Lagrangian contains an infinite number of terms, however, this infinite series of terms can be resummed for the self-accelerating solutions.  

\subsection{The potential in the decoupling limit}

In order to derive a Lagrangian for the decoupling limit of the action (\ref{genlag}), we write $M_{Pl}$ in terms of $m$ using (\ref{dec_lim}), and then take the $m\rightarrow 0$ limit. To avoid to be overburdened with indexes, it is convenient to write tensors in a matrix form;  a given tensor $\cal B_{\mu \nu}$ may be written as $\cal B$. Using this matrix notation, 
the tensor $H_{\mu\nu}\equiv H$ reads
\be
H = g-\left({\mathbf 1}-\Pi\, \eta^{-1} - d A\, \im\right)\, \eta\, \left({\bf 1}-\Pi\, \im - d A\, \im\right)^T\,,
\ee
where $g= g_{\mu\nu}$, $\eta=\eta_{\mu\nu}$, $\eta^{-1} = \eta^{\mu\nu}$,
 $\Pi=\Pi_{\mu\nu}$, $d A= \partial_\mu A_\nu$, and $\bf 1$ is the identity matrix.

\smallskip

The quantity we are interested in is  ${\cal K}^{\mu}_{\,\,\nu}\,=\,g^{-1} {\cal K} \,=\,{\bf 1}-\sqrt{{\bf 1}-g^{-1}\,H}$; 
so it is more convenient to write ${\bf 1}-g^{-1} H$ as
\bea
{\bf 1}-g^{-1} H &=& g^{-1}\,\left({\bf 1}-\Pi\, \im - d A\, \im\right)\,\left(\eta-\Pi  - d A^T\right)\,
\\
&=&
\left({\bf 1}-m^2\,  h\,\im\right)
 \left\{
 \left(\bu-\eta   \Pi\right)^2
 -m\,\im d  A \,\left(\bu-\eta  \Pi\right)-m\,\left(\bu-\eta  \Pi\right)\,\im d  A^T+m^2\,
 \im d  A \,\im d  A^T \right\},
\nonumber
\eea
where in the last step we have written everything in terms of the canonically normalized
fields (\ref{cannorm}), and expand the expression up to second order in $m$.  There are also factors of the strong coupling scale $\Lambda_3$ defined in Eq. (\ref{dec_lim}) that we set to one. However, it is straightforward to put it back by means of dimensional analysis.

We define
\bea
P&\equiv&\left({\bf 1}-\im\,  \Pi \right),\\
L_1&\equiv&
\im \, d   A \,P+P\,\im\,  d  A^T,
\\
L_2&\equiv&
\im\, d  A \,\im
 d  A^T
\eea
so that we can succinctly write 
\bea
{\bf 1}-g^{-1} H &=&\left({\bf 1}-m^2\, h\,\eta\right)
 \left\{P^2\
  -m L_1+m^2 L_2 \right\} .
\eea
We can neglect tensor fluctuations since, in the decoupling limit, they only couple to the scalar but not to the vector mode. The scalar-tensor couplings have been already completely
classified in \cite{deRham:2010ik}, so we can use these results to complete the full Lagrangian. Therefore, from now on we set $h=0$.

\bigskip

The potential for the scalar and vector fluctuations we are interested in is given by (\ref{potentialU}). Without loss of generality, we focus on the $\alpha_3=\alpha_4=0$ case first, and then generalise the result to non-vanishing $\alpha_3$ or $\alpha_4$ in Appendix \ref{AppC}. For $\alpha_3=\alpha_4=0$, the potential (\ref{potentialU}) simply reads
($\langle \cal K \rangle\,\equiv\,\tr {\cal K}$)
\bea\label{potU2}
U &=& m^2U_2= M_{Pl}^2 \,m^2\,\sqrt{-g}\,\left( \langle {\cal K}^2 \rangle
- \langle {\cal K} \rangle^2
\right)\nonumber\\
&=& \frac{\sqrt{-g}}{m^2}\,\left( \langle {\cal K}^2 \rangle
- \langle {\cal K} \rangle^2
\right)
\eea
where, again, we set $\Lambda_3  = 1$. Calling $\left(\bu-g^{-1}\,H\right)\,=\,M$ so that
$g^{-1} {\cal K}\,=\, \bu -\sqrt{M}$, we obtain
\be
U \,=\, \frac{1}{m^2}\,\left[
 6 \langle \sqrt{M} \rangle + \langle M \rangle-\langle \sqrt{M} \rangle^2-12
 \right].
\ee\label{pot_def}
Thus all the problem is to calculate $ \tr  \sqrt{M} $, to second order in $m$. This will be enough since the potential (\ref{potU2}) has an overall factor of $m^{-2}$ and contributions appearing with powers of $m$ bigger than two will go to zero in the decoupling limit. Therefore, up to second order in $m$, we can write
\bea\label{Mdef}
M&=&P^2
 \left[{\bf 1}
  -m P^{-2}\,L_1+m^2 P^{-2}\,L_2 \right]\,, \nonumber \\
  &=&
  P^2
 \left[{\bf 1}
  -\frac{m}{2} P^{-2}\,L_1+\frac{m^2}{2} P^{-2}\,L_2 - \frac{m^2}{8} P^{-2}\,L_1\,P^{-2}\,L_1\right]^2\,,\nonumber \\
   &\equiv&
  P^2
 \left[{\bf 1}
  - m \,Q_1 +
m^2\,Q_2
\right]^2\,,
\eea
with
\bea
Q_1&\equiv& \frac{1}{2} P^{-2}\,L_1\,,\\
Q_2&\equiv& \frac{1}{2} P^{-2}\,L_2 - \frac{1}{4} P^{-2}\,L_1\,P^{-2}\,L_1\,.
\eea
Thus, the square root of $M$ can be written, up to ${\cal O}(m^2)$, as
\be\label{expsq}
\sqrt{M}\,=\,P\,\left[ {\bf 1}
  - m \,Q_1 +
m^2\,Q_2 \right]+m\,D+m^2 E,
\ee
where the matrices $D$ and $E$ are needed since $P$ and $1- m \,Q_1 +m^2\,Q_2 $ do not commute; $D$ and $E$ obey the equations determined by requiring that the square of equation (\ref{expsq}) reduces to (\ref{Mdef}), to second order in $m$.  Using (\ref{expsq}), one arrives to the following expression for the potential $U$,
 \bea
 U&=&4 \,\tr\left[ P Q_2+E \right] +\tr L_2 +{\cal O}(m)\\
 &=& -\frac14 \tr{\left[ P^{-1}\,\im d  A \,\im d  A+
 P^{-1}\,\im d  A^{T} \,\im d  A^T-
  P^{-1}\,\im d  A \,\im d  A^T
 +\im d  A^{T}\, P^{-2} \,\im d  A\, P
  \right]}\nonumber\\
  &&+\tr\left[{\im d  A \,\im d  A^T}\right] +4\,\tr E +{\cal O}(m),\label{simpotf}
 \eea
where we have neglected total derivatives and used that $\tr D =0$, which we will prove  later. Notice that all the quantities proportional to the inverse powers of $m$ disappear, and when taking the decoupling limit the ${\cal O}(m)$ term vanishes. However, we still need to determine $\tr E$. As mentioned before, by taking the square on both sides of Eq.~(\ref{expsq}), we get the following condition to second order in $m$,
\bea\label{cond2sat}
0&=&-m\, P\,\left[ \left(  Q_1-m Q_2\right),\,P\right]\,\left({\bf 1} -m Q_1\right)
+ m \left\{ D, P \left({\bf 1} - m Q_1 \right) \right\} \nonumber\\
&&+m^2 \left\{ P,\,E\right\} +m^2 D^2.
\eea
To first order in $m$, we find a matrix equation for $D$ given by
\bea\label{probeq}
 \left\{ D, \,P \right\}
&=&P  \left[ Q_1,\,P
\right],
\eea
that can be recast as
\be\label{eqD1}
D+ P^{-1} \,D \,P\,=\,
\left[Q_1,P\right].
\ee
The previous equation implies that $\tr D\,=\,0$, as promised.
In terms of the original vector and scalar fields, equation (\ref{eqD1})
can be rewritten as 
\bea\label{eqfD}
D+ P  \,D \,P^{-1}&=&\frac12 \left(P^{-1}\,\im d { A} \,P- \im d { A}
 \right)\,-\,\frac12 \left(P\,\im d { A}^T \,P^{-1}- \im d { A}^T \right).
\eea
In certain cases, in which $P$ and $A$ have particularly simple forms, a solution to the previous matrix equation can be guessed by inspection of the structure of the matrices. On the other hand, it is convenient to have a systematic method to deal with solutions of the previous equation. It is easy to check that a formal, general solution can be written
in terms of the following combinations of infinite series
\bea\label{solDgen}
2\,D &=& \sum_{n=0}^{\infty} a_n\,P^{-n} \,\im d A \, P^n +  \sum_{n=0}^{\infty} b_n\,P^{n} \,\im d A \, P^{-n} \nonumber
\\&
+&\sum_{n=0}^{\infty} c_n\,P^{-n} \,\im d A^T \, P^n +  \sum_{n=0}^{\infty} d_n\,P^{n} \,\im d A^T \, P^{-n},
\eea
where the coefficients in the series satisfy the following relations
\begin{equation*}
\left\{\begin{array}{rl} a_n+a_{n+1}\,=\,0 &\text{ for  } n\ge 2\,,\\
b_n+b_{n-1}\,=\,0 &\text{ for  } n\ge 2\,,\\
a_0+b_0+b_1\,=\,0&\,\\
a_0+b_0+a_1\,=\,-1&\,\\
a_1+a_2\,=\,1&
\end{array}
\right.
\hskip1cm \text{and} \hskip1cm
\left\{\begin{array}{rl} c_n+c_{n+1}\,=\,0 &\text{ for  } n\ge 2\,,\\
d_n+d_{n-1}\,=\,0 &\text{ for  } n\ge 2\,,\\
c_0+c_1+d_0\,=\,1&\,\\
c_0+d_0+d_1\,=\,-1&\,\\
c_1+c_2\,=\,0&
\end{array}
\right.
\end{equation*}
 Notice that
the previous relations can be satisfied in different ways. For example, setting the coefficients
$b_n=0=d_n$, one obtains
\bea\label{solD}
2\,D &=&-P^{-1}\,\im d A\, P+2\,\sum_{n=2}^{\infty}\,\left(-1\right)^n\,P^{-n} \,\im d A\, P^n\nonumber\\
&& +\, \im d A^T +2 \,
 \sum_{n=1}^{\infty}\,\left(-1\right)^n\,P^{n} \,\im d A^T \,P^{-n}\,.
\eea
We will later discuss a method that allows to resum these series, but for the moment,
let us continue determining the quantity $\tr E$, that is needed to calculate
the potential in Eq.~(\ref{simpotf}). From Eq.~(\ref{cond2sat}), to second order in $m$, one finds
\bea\label{eqD2}
0&=&  P\,\left[  Q_2 ,\,P\right]+
 P\,\left[   Q_1,\,P\right]\, Q_1
- \left\{ D, P   Q_1  \right\}
+ \left\{ P,\,E\right\} + D^2 ,
\eea
which after some manipulations and using (\ref{eqD1}),
 (\ref{eqD2}) can be put into the form
\be\label{eqq2}
0\,=\,
\left[  Q_2 ,\,P\right]+\left[ D,\,Q_1\right]+ P^{-1}\,\left\{ P,\,E\right\} +P^{-1}\, D^2 .
\ee
Taking the trace, one finds
\be\label{trofe}
\tr{E }\,=\,-\frac12\,\tr\left( P^{-1}\, D^2 \right).
\ee
In order to get $\tr E$, the only unknown quantity\footnote{As we will see
in  section \ref{sec-a3a4}, in case where $\alpha_3$ or $\alpha_4$ do not vanish, we need the complete expression for $E$, and not only its trace, since the potential depends on $M^{3/2}$ as well as $\sqrt{M}$.} in (\ref{simpotf}),
one needs to determine $\tr\left( P^{-1}\, D^2 \right)$. Plugging the solution for $D$ of Eq.~(\ref{solD}), the previous quantity involves the resummation of an infinite series. Therefore, we learned a very interesting fact: even in the decoupling limit, the Lagrangian describing vector degrees of freedom contains an infinite number of terms. This has to be compared with the case of scalars only, in which we only have a finite number of terms corresponding to the Galileon combinations \cite{deRham:2010ik}.

\subsection{A method for resuming the series}

We  now  discuss a method, valid in many physically 
interesting cases, for resuming  the infinite series we encountered in the previous subsection. This method aims to rewrite those series in terms of geometric series. These can be easily resumed, leading to the non-polynomial
vector-scalar 
 couplings that we met in Section 3.2.

\smallskip
We  make the hypothesis that the matrix $P$ is diagonalisable. We write it as
\be\label{decfp}
 P \,= \,U \Upsilon U^{-1},
\ee
where $U$ is a unitary matrix, while $\Upsilon$ is a diagonal matrix that has the eigenvalues $\upsilon_i$ of $P$ in the diagonal. Eigenvalues and eigenvectors of $P$ are particularly simple
to calculate in the spherically symmetric case, since the matrix $P$ splits
into $2 \times 2$ matrices with particularly simple structure. For more general cases, extra work is needed.

As seen in (\ref{solDgen}), we would like to calculate the quantities
\be
P^{-n}\, \eta^{-1} d A \,P^n, \hskip1cm P^{n}\, \eta^{-1} d A \,P^{-n},
\hskip1cm
P^{-n}\, \eta^{-1} d A^T \,P^n, \hskip1cm P^{n}\, \eta^{-1} d A^T \,P^{-n}.
\ee
Considering for example the first of the previous quantities,
we can write
\be
P^{-n}\, \eta^{-1} d A \,P^n\,=\,U\,\Upsilon^{-n}\,B\,\Upsilon^n\,U^{-1},
\ee
with
$$
B\,\equiv\,U^{-1}\, \eta^{-1} d A \,U\,\equiv\,\left\{ b_{p q} \right\}_{M},
$$
where $\left\{ b_{p q} \right\}_{M}$ indicates a matrix with a component $b_{pq}$ at the $p-$th row and the $q$-th column. Then, we use the following identity
\be
\Upsilon^{-n}\,B\,\Upsilon^n\,=\, B+  \left\{ \left[\sum_{k=0}^{n-1}\left(\frac{\upsilon_q}{\upsilon_p}\right)^k\right]
\left( \frac{\upsilon_q}{\upsilon_p}-1\right)\,b_{pq}
 \right\}_M.
\ee
The case $n=1$ is easy to check, and for general $n$, it can be proved by induction.
The geometric series in the previous formula can be then resumed, thus one gets \bea
\Upsilon^{-n}\,B\,\Upsilon^n&=& B+  \left\{
\left[\left( \frac{\upsilon_q}{\upsilon_p}\right)^n-1\right]\,b_{pq}
\right\}_M\\
&=&
 \left\{
 \left( \frac{\upsilon_q}{\upsilon_p}\right)^n\,b_{pq}
\right\}_M.
\eea
This is what we need to resum explicitly the series appearing in the expression
 (\ref{solDgen}), which reduces to a combination of geometric series.

To conclude, once the eigenvalues and eigenvectors of $P$ are known, our method allows to rewrite the infinite series contained in $\tr E$ in terms of geometric series
that can be straightforwardly resumed.  Instead of providing general but lengthy formulae, we will discuss concrete and interesting examples in what follows. 

\subsection{An example}
Let us apply the previous method to a representative example.
We adopt spherical symmetry, and write a particular Ansatz which describes 
the configuration as in (\ref{diffeocanon}). Thus, we consider the following setup
\bea\label{newans}
 A_\mu&=&( A_t(r),0,0,0),\nonumber \\
 \pi&=& \pi_t(t)+ \pi_r(r).
\eea
Then
\begin{equation}
P = \left(
\begin{array}{cccc} 1+\ddot{\pi}& 0 & 0&0 \\ 0  & 1-\pi'' & 0&0 \\0&0&1-\pi'/r&0\\0&0&0&1-\pi'/r
\end{array} \right) ,\qquad
\im d A = \left(
\begin{array}{cccc} 0 & -A_t'(r) & 0&0 \\ 0  & 0 & 0&0 \\0&0&0&0\\0&0&0&0
\end{array} \right) .
\end{equation}
Using the formulae introduced in the previous section,
it is straightforward to check that
\bea
P^{-n}\, \im d A\,P^{n}&=&\left( \frac{1-\pi''}{1+\ddot{\pi}}\right)^n\, \im d A\, ,\\
P^{n}\, \im d A^T\,P^{-n}&=&\left( \frac{1-\pi''}{1+\ddot{\pi}}\right)^n\, \im d A^T.
\eea
Applying these results to the solution for $D$ written in Eq.~(\ref{solD}),
we can easily see that those series become the geometric series. We can indeed write
\bea\label{solDex}
2\,D &=&-P^{-1}\,\im d A\, P
+ \im d A^T
+2\left( \im d A + \im d A^T \right)\,\sum_{n=2}^{\infty}\,
\left( \frac{\pi''-1}{1+\ddot{\pi}}\right)^n.
\eea
The series is convergent if $|(1-\pi'')/(1+\ddot{\pi})|\,<\,1$, a condition that, a posteriori, can be checked to be satisfied for our selfaccelerating configuration (\ref{diffeocanon}). Using the usual formula for resuming the geometric series, we get
the explicit solution for $D$ as 
\be
D\,=\,\frac{(\pi''-1)\,(\ddot{\pi}+\pi'')}{2 \,(1+\ddot{\pi}) (2+\ddot{\pi} -\pi'')}\,\im d A+ \frac{ (\ddot{\pi}+\pi'')}{2 \,
(2+\ddot{\pi} -\pi'')}\,\im d A^T.
\ee
Substituting the solution for $D$ in the expression for the trace of $E$, Eq.~(\ref{trofe}), one recovers the scalar-vector coupling in the Lagrangian (\ref{dec_lag}). The appearance of scalar derivatives at the denominator of the expression is now easy to understand; it is  due to the resummation of an infinite number of terms in the geometric series.

\smallskip

Before moving on to the stability of the self-accelerating solution with a vector charge, we consider at this stage a more general Ansatz preserving the spherical symmetry, which will be necessary for the perturbation analysis. Consider $$\pi =  \pi(t,r) \qquad\mathrm{and}\qquad  A_{\mu} = ( A_t(t,r), A_r(t,r), 0, 0)$$ as Ansatz for the scalar and vector that preserves the spherical symmetry. Then the resulting interaction term between the scalar and the vector, calculated using the previous resummation techniques, is given by the following contribution to the action
 \be
S_{SV}=\int r^2 dtdr \Bigg\{\frac{2 (1+\ddot{\pi}-\pi''-2\frac{\pi'}{r}) \dot{A}_r
   A_t'+(1+2 \frac{\pi'}{r}) \left(\dot{A}_r^2+
   A_t'^2\right)}{2 (2-\pi''+\ddot{\pi})}
   -A_r'\dot{A}_t+\frac{2}{r} A_r \left(A_r'-\dot{A}_t\right)+
   \frac{1}{r^2} A_r^2\Bigg\},
\ee
which after integration by parts and using the divergence-free condition for the vector ($(r^2 A_r)'=r^2\dot{A}_t$), takes the simpler form
\be\label{Ssv}
S_{SV}
= \frac12\,\int
\frac{r^2\,dt dr }{ (2 + \ddot{\pi}-\pi'')}\,\left(1+\frac{2 \pi'}{r} \right)\,
\left( A_{t}'-\dot{A}_r\right)^2.
\ee

\section{Stability of the self-accelerating  configuration ($\alpha_3=\alpha_4=0$ case)}\label{secstabsim}
In this section, we analyse the perturbative stability of the self-accelerating configuration (\ref{vectorsol}), in the simple case of $\alpha_3=\alpha_4=0$; we will discuss the most general case in the following section. For simplicity, we consider perturbations that preserve the spherical symmetry. We discuss their dynamics, and investigate the existence of ghost degree of freedoms around our configurations.
We separate the discussions of this section in different subsections, which focus on different contributions to the total Lagrangian. In the last subsection we collect the results and study the sign of the kinetic terms for the relevant modes. We show that 
a ghost is always present around the self-accelerating configuration with a background vector field.

\subsection{Scalar-Vector Lagrangian}
We can decompose the divergence-free vector $A_\mu$ in terms of scalar and vector components with respect to the 2-sphere. Vector components are free and do not couple to $\pi$, so we can neglect their dynamics while studying the stability of the background. Thus in the following we only consider the scalar perturbations with respect to the 2-sphere. The residual U(1) gauge freedom of Eq.~(\ref{diffeodeco}) allows us to choose a gauge for $A_{\mu}$ so that it has only time and radial components. As a result, the most general Ansatz for scalar and vector field that preserves spherical symmetry for our system is
\begin{equation}\label{ansatzspherical}
\pi =  \pi(t,r), \quad \qquad
A_{\mu} = ( A_t(t,r), A_r(t,r), 0, 0).\,
\end{equation}
The action controlling interactions among the scalar $\pi$ and vector $A_{\mu}$ in the decoupling limit, is given by (\ref{Ssv}). We now consider the spherically symmetric, time-dependent perturbations around our self-accelerating configuration (\ref{solwq0}), namely
\begin{equation}
\pi = \pi_0 + \delta \pi(t,r), \quad
A_{\mu} = (A_{t0}(r) + \delta A_t(t,r), \delta A_r(t,r), 0, 0).
\end{equation}
where the background is given by 
\bea
\pi_0 &=& -\frac{1}{4} (r^2 -t^2) + \frac{3}{4}
\left(\sqrt{1+ \frac{16 Q_0^2}{9}} -1 \right) t^2, \\
A_{t0} &=& \frac{Q_0}{2} r^2.
\end{eqnarray}
The action (\ref{Ssv}), expanded at second order in perturbations, can be split as
\begin{equation}
S_{SV} = S_{SV}^{(1)} + S_{SV}^{(2)}
\end{equation}
where
\begin{equation}
S_{SV}^{(1)}
=\int dt dr r^2 \Bigg[ \frac{r}{8}
\Big(-3 + \sqrt{9+16 Q_0^2} \Big) \delta \pi'
\Bigg]
\end{equation}
is linear in the perturbations, and
\begin{equation}
S_{SV}^{(2)} = S^{S}_{SV} + S^{int}_{SV}\,, 
\end{equation}
\begin{eqnarray}
S^S_{SV} &=&
\int dt dr r^2 \Bigg [
\frac{4 Q_0^2 r }{(3+ \sqrt{9 + 16 Q_0^2})^2} \delta \pi' (- \delta \ddot{\pi} + \delta \pi'')
\Bigg], \\
S^{int}_{SV}
&=&\int dt dr r^2
\Bigg[
\frac{4 Q_0^2}{3+ \sqrt{9 + 16 Q_0^2}}\delta  \pi' ( \delta A_t' - \delta \dot{A}_r)
\Bigg],
\end{eqnarray}
is quadratic in the fluctuations.
Using the divergent free condition of the vector
\begin{equation}
- \delta \dot{A}_t +\delta A_r' + \frac{2 \delta A_r}{r} =0,
\end{equation}
we can show that $S^{(2)}_{SV}$
can be re-expressed  as
\begin{equation}\label{pertbaction}
S_{SV}^{(2)}=
\int dt dr r^2
\Bigg[
\frac{4 Q_0}{(3+ \sqrt{9 + 16 Q_0^2})^2}
\left( -\frac{3}{2} Q_0 (\delta \dot{\pi}^2 + \delta \pi^{'2})
+  (3+ \sqrt{9 + 16 Q_0^2}) (\delta \pi' \delta A_t' -\delta \dot{\pi} \delta \dot{A}_t)
\right)
\Bigg].
\end{equation}
As noticed before \cite{ArkaniHamed:2002sp}, if no background vector field is present there are no dynamical vector perturbations at the quadratic level. However, any small non-vanishing $Q_0$, produces dynamics in the vector degrees of freedom. Notice also that a non-vanishing $Q_0$ spontaneously breaks Lorentz invariance of the perturbed action in the scalar sector.

\subsection{Scalar-tensor Lagrangian}
In the decoupling limit, the scalar-tensor action is given by \cite{deRham:2010ik}
\begin{equation}
S_{ST}= \int dt dr r^2 \Bigg[
-\frac{1}{4} h^{\mu \nu} ({\cal E} h)_{\mu \nu} + h^{\mu}_{\,\,\nu} X^{\nu}_{\,\,\mu}
\Bigg],
\end{equation}
where
\begin{equation}
({\cal E} h)_{\mu \nu} = -\frac{1}{2}(\Box h_{\mu \nu} - \partial_{\mu} \partial_{\alpha} h^{\alpha}_{\nu} - \partial_{\nu} \partial_{\alpha} h^{\alpha}_{\mu} + \partial_{\mu} \partial_{\nu} h - \eta_{\mu \nu} \Box h + \eta_{\mu \nu}
\partial_{\alpha} \partial_{\beta} h^{\alpha \beta}),
\end{equation}
and
\begin{equation}\label{Xdef}
X^{\mu}_{\;\; \nu} =\frac{1}{2} \Big[\Pi \delta^{\mu}_{\;\; \nu} - \Pi^{\mu}_{\;\; \nu}
+\Pi^{\mu}_{\;\; \alpha} \Pi^{\alpha}_{\;\; \nu} - \Pi \,\Pi^{\mu}_{\;\; \nu}
+\frac{1}{2} (\Pi^2 -\Pi^{\alpha}_{\;\; \beta} \Pi^{\beta}_{\;\; \alpha}) \delta^{\mu}_{\;\; \nu}
\Big].
\end{equation}
Also in this case,  we consider perturbations around our self-accelerating configuration,
\begin{equation}
h_{\mu \nu}= f(t,r)\, \eta_{\mu \nu} + \delta \chi_{\mu \nu},
\quad f(t,r)= - \frac{1}{8} (r^2 -t^2),
\end{equation}
so that the scalar-tensor action becomes
\begin{equation}
S_{ST}=S_{ST}^{(1)} + S_{ST}^{(2)},
\end{equation}
where
\begin{equation}
S_{ST}^{(1)} = \int dt dr r^2 \Bigg[
f(t,r)\, X^{(1) \mu}_{\;\; \mu} - \frac{1}{2} \delta \chi^{\mu \nu}
\left[{\cal E}\, f(t,r)\, \eta\right]_{\mu \nu} +\delta  \chi^{\mu}_{\;\; \nu} X^{(0) \nu}_{\;\;\mu}
\Bigg],
\end{equation}
with 
\begin{eqnarray}
S_{ST}^{(2)} &=& S_{ST}^S + S_{ST}^{int} +S_{ST}^T,\\
S_{ST}^S &=& f(t,r)\, X^{(2) \mu}_{\;\; \mu}, \\
S_{ST}^{int} &=& \delta \chi^{\mu}_{\;\;\nu} X^{(1) \nu}_{\;\; \mu}, \\
S_{ST}^T &=&  - \frac{1}{4} \delta \chi^{\mu \nu} ({\cal E} \delta \chi)_{\mu \nu},
\end{eqnarray}
and $X^{(n)}$ the $n^{\mathrm{th}}$ perturbation of (\ref{Xdef}).
Given our field configurations, the non-vanishing components of the tensors $X^{(i)}$ read
\begin{eqnarray}
X^{(0) \mu}_{ \,\,\,\,\;\;\;\;\nu} &=& -\frac{3}{8} \delta^{\mu}_{\;\; \nu},\\
X^{(1) \mu}_{\,\,\,\,\;\;\;\; \mu} &=& -\frac{1}{2}
\Big(-3 +\sqrt{9+16 Q_0^2} \Big)
\Bigg( \delta \pi'' + 2 \frac{\delta \pi'}{r} \Bigg),\\
X^{(2) \mu}_{\,\,\,\,\;\;\;\; \mu}
&=& \delta \dot{\pi}'^2 + \frac{\delta \pi'}{r}
\Bigg(2 \delta \pi'' + \frac{\delta \pi'}{r} \Bigg)
-\delta \ddot{\pi} \Bigg(\delta \pi'' + \frac{2 \delta \pi'}{r} \Bigg).
\end{eqnarray}

\smallskip

Starting from the previous expressions, we can show that
\begin{equation}
S_{SV}^{(1)} + S_{ST}^{(1)} =0,
\end{equation}
as expected, since first order perturbations should vanish using the equations of motion for the background. Furthermore, the pure scalar field perturbation simplifies to
\begin{equation}
S_{ST}^S = \int r^2\,dt dr  \Bigg[
-\frac{3}{8} (\delta \dot{\pi}^2 - \delta \pi'^2)
\Bigg].
\end{equation}

Finally, we only need to focus on the scalar part of the tensor perturbations with respect
to the 2 sphere, since it is the one that has non-trivial
dynamics due to the coupling with the scalar degrees of freedom. It is
convenient to use the gauge \cite{soda} 
\begin{equation}\label{xidec}
\chi^{\mu}_{\,\,\nu} = \left(
\begin{array}{cccc} -H_t & - H_1 & 0 &0  \\ H_1  & H_r  & 0 & 0  \\ 0 & 0 & 0 &0\\
 0 & 0 & 0 &0
\end{array} \right).
\end{equation}
The tensor-scalar coupling is then given by
\begin{equation}
S_{ST}^{int} = \int r^2\,dt dr \, \Bigg[
-\frac{1}{2 r} \Big(-3 + \sqrt{9+16 Q_0^2} \Big)\delta  \pi' H_r
\Bigg],
\end{equation}
while the Einstein-Hilbert action reduces to
\begin{equation}
S^T_{ST} = \int dr dt r^2 \Bigg[ \frac{1}{2 r^2}
\left( 2 r \dot{H}_r H_1 + \frac{1}{2} H_r^2 - \frac{1}{2} H_t (2 r H_r)'
\right)
\Bigg].
\end{equation}
By taking variations with respect to $H_1$ and $H_t$, we find that $H_r$ is not dynamical. 
Thus, tensor degrees of freedom do not play any role in the dynamics of the system in the spherically symmetric case, and their dynamics can be consistently neglected.

\subsection{The kinetic terms, and the inevitable ghost degree of freedom}
Collecting the previous results  we obtain the second order action for the scalar and vector field perturbations, given by
\begin{eqnarray}
S^{(2)} =
\int r^2 dt dr
\Bigg[
-\left(\frac{3}{8} + \frac{6 Q_0^2}{(3+ \sqrt{9 + 16 Q_0^2})^2}
\right) \delta \dot{\pi}^2
+ \left(\frac{3}{8} -  \frac{6 Q_0^2}{(3+ \sqrt{9 + 16 Q_0^2})^2}
\right) \delta \pi'^2 \nonumber\\
+ \frac{4 Q_0}{3+ \sqrt{9 + 16 Q_0^2}}(\delta \pi' \delta A_t' -\delta \dot{\pi} \delta \dot{A}_t)
\Bigg]\,.
\end{eqnarray}
It is straightforward to diagonalize the kinetic terms, and identify ghost-like degrees of freedom. Defining
\begin{equation}
\alpha = \frac{3}{8} + \frac{6 Q_0^2}{(3+ \sqrt{9 + 16 Q_0^2})^2}, \quad
\beta = \frac{2 Q_0}{3+ \sqrt{9 + 16 Q_0^2}},
\end{equation}
we get
\begin{equation}
S^{(2)} = \int dt dr r^2 \Bigg[
(\delta \pi, \,\delta  A_t)
\left(
\begin{array}{cccc} \alpha & \beta    \\ \beta  & 0
\end{array} \right)
\frac{d^2}{dt^2}
\left(
\begin{array}{c} \delta \pi    \\ \delta A_t
\end{array} \right)
 \Bigg] + ...
\end{equation}
Then by using the the following definitions
\begin{eqnarray}
\delta \tilde{\pi} &=& \delta \pi + \delta A_t, \\
\delta \tilde{A}_t &=& \frac{\alpha - \sqrt{\alpha^2 + 4 \beta^2} }{2 \beta} \delta \pi
+ \frac{\alpha + \sqrt{\alpha^2 + 4 \beta^2} }{2 \beta} \delta A_t,
\end{eqnarray}
the kinetic term reads
\begin{equation}\label{kinaction}
S^{(2)} = \int dt dr r^2 \Bigg[
(\delta \tilde{\pi}, \delta \tilde{A}_t)
\left(
\begin{array}{cccc} \lambda_1 & 0    \\ 0  & \lambda_2
\end{array} \right)
\frac{d^2}{dt^2}
\left(
\begin{array}{c} \delta \tilde{\pi}    \\ \delta \tilde{A}_t
\end{array} \right)
 \Bigg] + ...
\end{equation}
where
\begin{equation}
\lambda_1= \frac{1}{2} (\alpha + \sqrt{\alpha^2 + 4 \beta^2}),
\quad
\lambda_2 = \frac{1}{2} (\alpha - \sqrt{\alpha^2 + 4 \beta^2}).
\end{equation}
A ghost is present if one of the eigenvalues $\lambda_{i}$ is positive. It is easy to see
that $\lambda_1 \geq 3/8$ and $\lambda_2 \leq 0$, with both equalities satisfied when $Q_0=0$. 
Thus $\delta\tilde{\pi}$ is {\it always} a ghost while  $\delta \tilde{A}_t$ is a normal mode, regardless of the size of the vector charge $Q_0$. Notice, however, that when $Q_0=0$ the kinetic term for the mode $\delta \tilde{A}_t$ vanishes and we enter in a regime
of strong coupling. The background vector field with $Q_0\neq 0$ then alleviates the strong coupling behaviour providing a healthy kinetic term for the vector fluctuations. In the next
section, we investigate whether these conclusions change when $\alpha_3$ and $\alpha_4$, and thus higher powers of ${\cal K}$, are included in the potential.

\section{The case of non-vanishing $\alpha_3$ and $\alpha_4$}\label{sec-a3a4}

In this section we would like to discuss what happens in the general case, when higher powers of ${\cal K}$ are included in the massive gravity potential (\ref{potentialU}). The inclusion of such terms changes only the coefficients in front of each contributions,
but not the general structure of the solutions and the Lagrangian in the decoupling limit.
However, the existence of a ghost does depend in a subtle and surprising way on these coefficients, as we will see in what follows.

\subsection{Lagrangian in the decoupling limit}\label{Lag_a3a4}

In order to construct such Lagrangian using the material developed in Section \ref{Lagsection}, we first write the potential (\ref{potentialU}) in terms of $M$, resulting in
\bea
U\!\! &=&\!\! - m^2\Bigg\{\left(\langle\sqrt{M}\rangle-6\right) \langle\sqrt{M}\rangle +
 \alpha_3 \left[24 - 2 \langle M^{3/2}\rangle + 3 \langle M\rangle \left(\langle\sqrt{M}\rangle-2\right) -
    \langle\sqrt{M}\rangle \left(18 + \left(\langle\sqrt{M}\rangle-6\right) \langle\sqrt{M}\rangle\right)\right] \nonumber \\ &&+
 \alpha_4 \bigg[3 \left(8 + \left( \langle M\rangle-4\right) \langle M\rangle - 8 \langle\sqrt{M}\rangle\right)
-6 \langle M^{2}\rangle + 8 \langle M^{3/2}\rangle \left(\langle\sqrt{M}\rangle-1\right) +
     \langle\sqrt{M}\rangle \Big(-6 \langle M\rangle \left(\langle\sqrt{M}\rangle-2\right) \nonumber
\\&& + \langle\sqrt{M}\rangle \left(12 + \left(-4 + \langle\sqrt{M}\rangle\right) \langle\sqrt{M}\rangle\right)\Big)\bigg]
+12 - \langle M\rangle\Bigg\}.
\eea

Notice, that not only the traces of $\sqrt{M}$ and $M$ are needed, as in the $\alpha_3=\alpha_4=0$ case, but also the trace of $M^{3/2}$. Therefore, it is necessary to solve completely for $E$ using equation (\ref{eqq2}), and not only for its trace, as we did in section \ref{Lagsection}. In the case of the most general spherically symmetric Ansatz (\ref{ansatzspherical}), one gets a similar interaction Lagrangian to that of (\ref{Ssv}), but including terms with $\alpha_3$ and $\alpha_4$. Since its expression is long, we refer the reader to Appendix C for its explicit form. However, we can mention that the vector-scalar coupling is very similar and has the same expression in the denominator as before, {\it i.e.} $(2+\ddot{\pi}-\pi'')$. Therefore, the same concern about higher derivatives applies; on the other hand, the generalised self-accelerating solution is such that higher derivative terms vanish when perturbing around it. In what follows we construct such self-accelerating solution.

\subsection{Self-accelerating solution}
\begin{figure}[htp!]
\begin{center}
\includegraphics[scale=.5,bb=0 0 379 403]{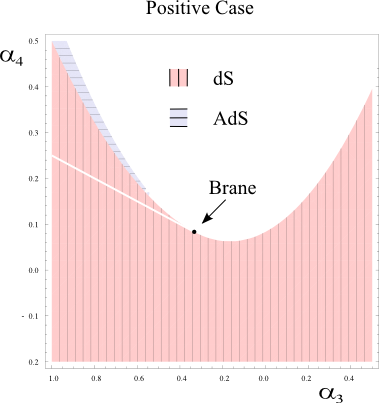} \hskip .5cm
\includegraphics[scale=.5,bb=0 0 377 403]{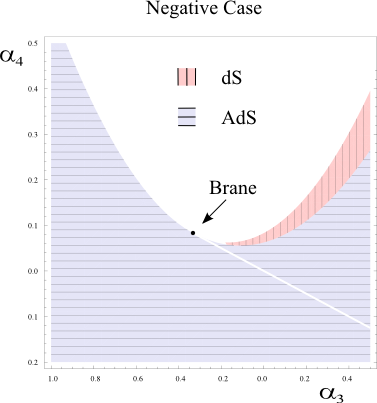}
\end{center}
\caption{The dS and AdS regions of the self-accelerating solution in the decoupling limit (\ref{decsola3a4}). Left (right) plot is for $\alpha_5>0$ ($\alpha_5<0$). The white exclusion area is associated  to the positiveness of the argument in the square root of $\alpha_5$. In each plot, half of the line $\alpha_3+3 \alpha_4=0$ has to be removed to avoid the curvature of the space-time from diverging, with the middle point being that of the brane action found in Ref.~\cite{us2}.
}
\label{fig1}
\end{figure}

The most general static spherically symmetric solution can be constructed in the same way as for the case studied in section \ref{sec-selfacc}. Again, for the full Lagrangian (\ref{genlag}), there are two branches of solutions: one with a diagonal metric and the other with an off-diagonal component. In contrast to the $\alpha_3=\alpha_4=0$ case, the diagonal branch presents perturbative solutions which do not decay at infinity, but we will not focus on them, since they do not recover general relativity, via the Vainshtein mechanism, for small radius \cite{us, us2, david, new}. The non-diagonal branch is very similar to previous case, and a full solution can be constructed with the same structure. Details are given in Appendix A. For the purposes of this paper, we only need the decoupling limit of this class of solutions, and which is given by (see Appendix B) 
\bea\label{decsola3a4}
h_{\mu\nu}&=&-\Lambda_3\left[\frac{(1 + 3 \alpha_3 + 2 \alpha_5)+(1 + 3 \alpha_3 + \alpha_5)^2\Lambda}{ 6(1 + 3 \alpha_3 + \alpha_5)^2}\right](r^2 - t^2)\eta_{\mu\nu},\nonumber\\
\pi&=&-\frac{1}{2}\left[1-\frac{(2+3 \alpha_3+\alpha_5)}{(1+3
   \alpha_3+\alpha_5)}\Gamma(\Lambda,Q_0,\alpha_3,\alpha_4)\right]\Lambda_3 t^2 -\frac{1}{2(1+3 \alpha_3+\alpha_5)}\Lambda_3 r^2,\nonumber \\
A_0&=& -\frac{Q_0}{2}\Lambda_3 r^2.
\eea
where
\be\label{gamma}
\Gamma(\Lambda,Q_0,\alpha_3,\alpha_4)\equiv\sqrt{1+\frac{3  (1+3 \alpha_3+\alpha_5)^4Q_0^2}{(2+3
   \alpha_3+\alpha_5)^2 [(1+3 \alpha_3+2 \alpha_5)+(1+3 \alpha_3+\alpha_5)^2\Lambda]}}.
\ee
We have assembled the parameters  $\alpha_3$ and $\alpha_4$ into a combination called $\alpha_5$, with
\be\alpha^2_5\equiv {1+3\alpha_3+9\alpha_3^2-12\alpha_4}\ee
to simplify considerably the expressions. Notice, that this solution includes the possibility of a bare cosmological constant $\Lambda$ in our formulae,
that we added for completeness, and using the normalisation (\ref{cannorm}). Moreover, $\alpha_5$ should be real for the solution to exist, resulting in 
\be
1+3\alpha_3+9\alpha_3^2-12\alpha_4 >0,
\ee
and it can have both signs, depending on the branch of the square root one considers. Figure \ref{fig1} shows the parameter space in which the solution (\ref{decsola3a4}) exists for both, positive and negative signs, of $\alpha_5$.
Therefore, the limit $\alpha_3=\alpha_4 = 0$ can only be taken in the positive square root branch, otherwise one is lead to singular expressions. The sign of the combination $\left[1 + 3 \alpha_3 + 2 \alpha_5+(1 + 3 \alpha_3 + \alpha_5)^2\Lambda
\right]$ sets whether the solution is de Sitter or anti-de Sitter, and one can associate an effective cosmological constant\footnote{In the expression for the effective cosmological constant we have not dropped the normalisation of (\ref{cannorm}).} given by 
\be 
\Lambda_{eff, \pm} = \Lambda + m^2 \frac{1 + 3 \alpha_3 \pm 2 \sqrt{1+3\alpha_3+9\alpha_3^2-12\alpha_4 }}{(1 + 3 \alpha_3 \pm \sqrt{1+3\alpha_3+9\alpha_3^2-12\alpha_4 })^2}.
\ee
In the AdS case, there is an extra constraint on $Q_0$ due to the argument in the square root of (\ref{gamma}), and given by 
\be
Q_0^2\,\le\,\frac{(2 + 3 \alpha_3 + \alpha_5)^2\Big|(1+3 \alpha_3+2 \alpha_5)+(1+3 \alpha_3+\alpha_5)^2\Lambda\Big|}{3(1 + 3 \alpha_3 + \alpha_5)^4}.
\ee

Although we derived these decoupling solutions from the exact spherically symmetric solution, other exact solutions in the full theory may have the same decoupling limit. For example, the open FRW solution found in Ref.~\cite{Gumrukcuoglu:2011ew} also reduces to Eq.~(\ref{decsola3a4}) with $Q_0=0$ in the decoupling limit. 

As expected, the previous expressions (\ref{decsola3a4}) solve the equations of motion obtained from the decoupling limit Lagrangian, which was discussed in the previous Section and whose full expression is given by (\ref{laga3a4}). Now, we would like to understand the stability of these, more general, self-accelerating solutions, which include vector degrees of freedom and a bare cosmological constant.

\subsection{Linear perturbations: choosing between a ghost or strong coupling}
We now study spherically symmetric perturbations around the background (\ref{decsola3a4}) 
as we did for the $\alpha_3=\alpha_4=0$ case in Section \ref{secstabsim}. The procedure is exactly the same, and we refer the reader to Appendix C where the explicit calculations are shown. Here, we only present the results. The Lagrangian controlling the perturbations is
\bea\label{actionperts}
S^{(2)}&=&\int r^2 dt dr \bigg\{\frac{\alpha_5 \left[1+3 \alpha_3+2 \alpha_5+ (1+3
   \alpha_3+\alpha_5)^2\Lambda\right]\left[\delta\pi'^2-\Gamma(\Lambda,Q_0,\alpha_3,\alpha_4)\delta\dot{\pi}
   ^2\right]}{(1+3
   \alpha_3+\alpha_5)^2[1+\Gamma(\Lambda,Q_0,\alpha_3,\alpha_4)]}
    \\ \nonumber  &&-\frac{2 \alpha_5  Q_0 (1 + 3 \alpha_3 + \alpha_5)}{(2+3
   \alpha_3+\alpha_5)[1+\Gamma(\Lambda,Q_0,\alpha_3,\alpha_4)]}
   \left(\delta A_t'\,  \delta\pi'
   -\delta \dot{A}_t \, \delta\dot{\pi}\right)\bigg\},
\eea
where $\Gamma$ is defined in (\ref{gamma}).
Therefore, we only need to consider fluctuations in $\delta \pi$, and in $\delta A_t$ for the vector. Their kinetic terms can be straightforwardly diagonalised, and the resulting action is again
  (\ref{kinaction}), but with eigenvalues $\lambda_{1,2}= \frac{1}{2} (\alpha + \sqrt{\alpha^2 \pm 4 \beta^2})$, where $\alpha$ and $\beta$ are now given by
\bea
\alpha&=&\frac{\alpha_5 \left[1 + 3 \alpha_3 +2 \alpha_5 + (1 + 3 \alpha_3 + \alpha_5)^2 \Lambda\right]\Gamma(\Lambda,Q_0,\alpha_3,\alpha_4)}{(1 + 3 \alpha_3 +\alpha_5)^2 [1+\Gamma(\Lambda,Q_0,\alpha_3,\alpha_4)]}\nonumber \\
\beta&=&-\frac{\alpha_5 \left([2 + 3 \alpha_3 + \alpha_5]^2\left[\Gamma(\Lambda,Q_0,\alpha_3,\alpha_4)^2-1\right]\left[1 + 3 \alpha_3 +2 \alpha_5 + (1 + 3 \alpha_3 + \alpha_5)^2 \Lambda\right]\right)^{1/2}}{\sqrt{3}(1 + 3 \alpha_3 + \alpha_5)[1+\Gamma(\Lambda,Q_0,\alpha_3,\alpha_4)]}.
\eea
In order to investigate the existence of ghost degrees of freedom, we have to study the sign of $\lambda_i$; if at least one of the two eigenvalues is positive, a ghost is present. In
order to study the ghost issue, it is convenient to focus on the product of the eigenvalues, whose expression reduces to
\be\label{prod}
\lambda_1\lambda_2=-\frac{\alpha_5^2 \left[1 + 3 \alpha_3 + 2 \alpha_5 + (1 + 3 \alpha_3 + \alpha_5)^2 \Lambda\right]\left[\Gamma(\Lambda,Q_0,\alpha_3,\alpha_4)-1\right]}{
 3 (1 + 3 \alpha_3 + \alpha_5)^2 [1 + \Gamma(\Lambda,Q_0,\alpha_3,\alpha_4)]}
\ee
It is simple to convince oneself that the previous product is always negative when $Q_0 \neq 0$; thus a ghost is {\it always} present around the self-accelerating solution with the non-vanishing vector field because one of the modes have kinetic terms with the wrong sign.  The result is independent on the values of $\alpha_i$ and $\Lambda$, as long as we have dS (or AdS) symmetry. On the other hand, the kinetic term for the other mode is positive definite and always healthy.

On the other hand, when $Q_0=0$ ($\Gamma=1$), the product (\ref{prod}) vanishes. The ghost can
be avoided but a strong coupling appears since the kinetic term for one of the fluctuations goes to zero in this limit. It is not the purpose of this paper
to investigate in detail the physical consequences of this strong coupling regime. Here, we are only interested in the conditions that avoid the ghost, generalizing
the analysis of \cite{deRham:2010tw} to the case of non-vanishing bare cosmological
constant. Therefore, when $Q_0=0$ the eigenvalues read
\be\label{eigeq0}
\lambda_{1,2}=\frac{1}{4 (1 + 3 \alpha_3 + \alpha_5)^2}\left[\alpha_5 (1 + 3 \alpha_3 + 2 \alpha_5 + (1 + 3 \alpha_3 + \alpha_5)^2 \Lambda)\pm |\alpha_5| |1 + 3 \alpha_3 + 2 \alpha_5 + (1 + 3 \alpha_3 + \alpha_5)^2 \Lambda|\right].
\ee
We assume the Hubble parameter squared is positive (de Sitter regime), which from (\ref{decsola3a4}) translates into the condition
\be
1 + 3 \alpha_3 + 2 \alpha_5 + (1 + 3 \alpha_3 + \alpha_5)^2 \Lambda>0.
\ee
Comparing this with Eq.~(\ref{eigeq0}), we conclude that the only way to avoid the ghost is to choose $\alpha_5<0$. To summarize, the conditions to avoid the ghost in a de Sitter regime are
\be
\Lambda > \frac{-1 - 3 \alpha_3 + 2 |\alpha_5|}{(1 + 3 \alpha_3 - |\alpha_5|)^2} \qquad {\rm and}\qquad \alpha_5<0\,\,.
\ee
These conditions generalize the results previously obtained in the literature, and 
when $\Lambda=0$ they reduce to those in \cite{deRham:2010tw}.

\section{Conclusions}

In this paper, we presented new self-accelerating solutions for a family of non-linear massive gravity models in the decoupling limit. The dynamics of vector degrees of freedom of the massive graviton is fully taken into account, playing an important role in characterizing these configurations. We found two branches of solutions, which describe a de-Sitter spacetime in different regions of the parameter space $(\alpha_3,\alpha_4)$. In order to study the dynamics of linear fluctuations, we developed the tools to obtain a general Lagrangian in the decoupling limit, which includes vector degrees of freedom. The resulting Lagrangian contains an infinite number of terms, which lead to non-polynomial interactions between the vector and scalar degrees of freedom.

We then studied linear fluctuations around these self-accelerating configurations, finding two different behaviours. When the background vector field is switched on, a ghost mode is always present in the spectrum of propagating modes, regardless of the choice of parameters characterizing the theory. In contrast, when the background vector field is set to zero, our configurations reduce to those analysed in  \cite{deRham:2010tw}. In this case, ghosts may be absent in one of the two branches of solutions and in some regions of parameter space. This ghost-free branch does not exist for the simplest case with $\alpha_3=\alpha_4=0$. However, the coefficient in front of the kinetic terms for the vector fluctuations vanishes, leading to a strong coupling effect.

A bare cosmological constant can be fully included in these self-accelerating configurations, and the same conclusions hold. There is ghost mode if a background vector field exists, and the ghost may be removed if this background field is not present. The avoidance of the ghost is again only in one of the two branches of solutions, and for a positive bare cosmological constant the parameter space with a ghost-free model is enlarged.

\smallskip

Our analysis can be extended in many interesting directions. It is important to investigate in more details the Lagrangian in the decoupling limit with vectors. 
It would be interesting to analyse how matter couples to the degrees of freedom controlled by this Lagrangian and understand the role of vectors. Regarding the self-accelerating solutions, it would also be interesting to understand whether symmetries exist that automatically set the background vector field to zero and avoid ghost modes. Imposing Lorentz invariance on the solution of the scalar mode, $\pi$, can achieve this goal; but a more comprehensive analysis of possible symmetries is needed. It is also crucial to clarify the physical consequence of the strong coupling regime in the vector fluctuations, that one encounters when the vector background profile is switched off. As stated in  \cite{deRham:2010tw}, quantum corrections might be able to induce a healthy kinetic
term for the vector; it would be interesting to understand this effect in more detail, and to discuss its
consequences for the self-accelerating configurations.
Finally, it is encouraging that there exist ghost-free self-accelerating solutions in the decoupling limit, but this does not necessarily ensure that the full non-linear solution is stable, especially when a bare cosmological constant is included. This last point may be relevant for inflationary cosmology within this theory of massive gravity. A detailed analysis of full non-linear self-accelerating solutions away from the decoupling limit will be presented in a future publication \cite{new2}.

\subsection*{Acknowledgements}
KK and GN are supported by  the European Research Council. KK also acknowledges support from the STFC grant ST/H002774/1 and the Leverhulme Trust.
GT is supported by an STFC Advanced Fellowship ST/H005498/1.

\begin{appendix}
\section{General exact solution}\label{AppA}

From the general Lagrangian (\ref{genlag}), and using the non-diagonal ansatz (\ref{genmetr}) together with Einstein equations (\ref{einstein}), one can show that there are two branches of solutions for non-vanishing $\alpha_3$ and $\alpha_4$. These solutions were studied in detail in \cite{us2}, however, here we only consider the branch with a non-diagonal metric, where analytic solutions can be found. Since the combination $\sqrt{1+3\alpha_3+9\alpha_3^2-12\alpha_4}$ is always present in the solution of this branch, it is convenient to map the ($\alpha_3,\alpha_4$) parameters into ($\alpha_3,\alpha_5$), where $\alpha_5^2\equiv1+3\alpha_3+9\alpha_3^2-12\alpha_4$. In this new set of parameters, the combination  $D(r)\, G_{tt}+C(r)\,G_{tr}\,=\,0$, fixes $B$ as a function of $r$ in the following way
\be\label{boapp}
B(r)= b_0 r^2 \,=\,
\frac{(1+3 \alpha_3+\alpha_5)^2}{(2+3 \alpha_3+\alpha_5)^2} r^2.
\ee
The rest of Einstein equations give
\be\label{generalsol}
C(r)= \frac{\Delta_0}{b_0}(1-p), \qquad A(r)=\frac{\Delta_0}{b_0}(p+\gamma+1), \qquad D(r)=\sqrt{\Delta_0-A(r)C(r)},
\ee
where
\be
p=\frac{c}{r}+\frac{(1+3 \alpha_3+2 \alpha_5)}{3 (2+3 \alpha_3+\alpha_5)^2}m^2 r^2, \qquad
\gamma+1=\frac{(1+3 \alpha_3+\alpha_5)^4}{\Delta_0(2+3 \alpha_3+\alpha_5)^4}
\ee
Just like in the $\alpha_3=\alpha_4=0$ ($\alpha_5=1$) case, there are two integration constants, $c$ and $\Delta_0$, but in order to have a positive argument for the square root in $D(r)$, $\Delta_0$ has to run from $\Delta_0=0$ to $\Delta_0^{max}=b_0^2$. In this paper, we focus on the massless case $c=0$ only, which describes the static patch of the de Sitter or Anti-de Sitter spacetime. 

%
%
%

\section{The decoupling limit}\label{AppB}
If one naively takes the decoupling limit (\ref{dec_lim}) of the static configuration (\ref{gmcoeff}) [or more generally of (\ref{boapp})-(\ref{generalsol})], one gets a divergent result. This is due to scalar and vector contributions which would be zero in this na\"{\i}ve limit. Therefore, in order to extract a finite answer, one should look for a coordinate transformation that, after expanding $h_{\mu\nu}=g_{\mu\nu}-\eta_{\mu\nu}$ in terms of $m$, leaves the first non-vanishing coefficient to be that of $m^2$, so that the limit $m\rightarrow 0$ is well defined.
For our case, this transformation is precisely given by 
\be\label{transf}
(t\ ,\ r)\rightarrow \left(\sqrt{\Delta_0 b_0} t+\int_0^r\frac{D(\rho)}{C(\rho)}d\rho\ ,\ \frac{r}{\sqrt{b_0}}\right),
\ee
where $b_0, D$ and $C$ are given in (\ref{gmcoeff}) [or more generally in (\ref{boapp})-(\ref{generalsol})]. The metric field $h_{\mu\nu}$, after being canonically normalised using the definition (\ref{cannorm}), reads
\be\label{hmunu1}
h_{\mu\nu}=\frac{(1 + 3 \alpha_3 + 2 \alpha_5)}{3 (1 + 3 \alpha_3 + \alpha_5)^2}r^2\eta_{\mu\nu}
\ee
Moreover, the transformation (\ref{transf}) induces the following components of the St\"uckelberg field $\pi^\mu$,
\bea
\pi^0&=&\left(\frac{1+3 \alpha_3+\alpha_5}{(2+3
   \alpha_3+\alpha_5)\sqrt{\Delta_0} }-1\right)t
   -\frac{1}{2}\sqrt{\frac{(1+3 \alpha_3+2 \alpha_5) \left[(1+3
   \alpha_3+\alpha_5)^4-(2+3 \alpha_3+\alpha_5)^4\Delta_0 \right]}{
   3\Delta_0 (1+3 \alpha_3+\alpha_5)^4 (2+3
   \alpha_3+\alpha_5)^2}}m r^2 +\mathcal{O}(m^2)\nonumber \\
\pi^r&=&-\frac{1}{1+3 \alpha_3+\alpha_5}r+\mathcal{O}(m^2),
\eea
which in turn, can be decompose into scalar and vector modes. After being canonically normalised using (\ref{cannorm}), the vector and scalar modes reduce to
\bea\label{vec-scalar}
\pi&=&-\left\{\frac{1}{2(1+3 \alpha_3+\alpha_5)}\left[r^2-(2+3 \alpha_3+\alpha_5)\sqrt{1+\frac{3(1+3 \alpha_3+\alpha_5)^4Q_0^2}{(2+3 \alpha_3+\alpha_5)^2(1+3 \alpha_3+2\alpha_5)}}t^2\right]+\frac{t^2}{2}\right\}\Lambda_3 
+\mathcal{O}(m), \nonumber \\
A_0&=& -\frac{Q_0}{2}\Lambda_3 r^2+\mathcal{O}(m),
\eea
where we have decided to measure the vector charge using $Q_0$ --- a real positive number --- instead of $\Delta_0$ (which was bounded from 0 to $b_0^2$). The following equation relates $Q_0$ and $\Delta_0$,
\be\label{q0delta0}
\Delta_0=\frac{(1+3 \alpha_3+\alpha_5)^4 (1+3 \alpha_3+2 \alpha_5)}{3
   Q_0^2 (2+3 \alpha_3+\alpha_5)^2 (1+3
   \alpha_3+\alpha_5)^4+(2+3 \alpha_3+\alpha_5)^4 (1+3
   \alpha_3+2 \alpha_5)}.
\ee
Finally, one can take the decoupling limit of expressions (\ref{hmunu1}) and (\ref{vec-scalar}), and get a well defined answer.

One can induce further coordinate transformations, of order $m^2$, which take the metric to any desired form in the decoupling limit and consequently do not affect the vector nor the scalar. Using this freedom, one can write the canonically normalised metric field $h_{\mu\nu}$ in a covariant conformal form, namely
\be
h_{\mu\nu}=-\left(\frac{(1 + 3 \alpha_3 + 2 \alpha_5) \Lambda_3(r^2 - t^2)}{6 (1 + 3 \alpha_3 + \alpha_5)^2}+\mathcal{O}(m)\right)\eta_{\mu\nu}.
\ee
Therefore for $(1 + 3 \alpha_3 + 2 \alpha_5)$ positive (negative) one gets de Sitter (Anti-de Sitter). In the case of AdS, one has an extra constraint on $Q_0$, from the square root argument in (\ref{vec-scalar}), given by
\be\label{adsbound}
Q_0^2<\frac{(2 + 3 \alpha_3 + \alpha_5)^2|1 + 3 \alpha_3 + 2 \alpha_5|}{3(1 + 3 \alpha_3 + \alpha_5)^4}.
\ee

One can include a bare cosmological constant, and still get a solution for the non-diagonal branch with a very similar form as (\ref{boapp})-(\ref{generalsol}). The details of this solution were described in \cite{us2}, but here we will only describe its decoupling limit. Therefore, the non-diagonal branch solution for the action (\ref{genlag}), which includes a cosmological constant $\Lambda$ (normalised as in (\ref{cannorm})), has the following decoupling limit
\bea
h_{\mu\nu}&=&-\Lambda_3\left[\frac{(1 + 3 \alpha_3 + 2 \alpha_5)+(1 + 3 \alpha_3 + \alpha_5)^2\Lambda}{ 6(1 + 3 \alpha_3 + \alpha_5)^2}\right](r^2 - t^2)\eta_{\mu\nu},\nonumber\\
\pi&=&-\frac{1}{2}\left(1-\frac{(2+3 \alpha_3+\alpha_5)}{(1+3
   \alpha_3+\alpha_5)}\sqrt{1+\frac{3  (1+3 \alpha_3+\alpha_5)^4Q_0^2}{(2+3
   \alpha_3+\alpha_5)^2 [(1+3 \alpha_3+2 \alpha_5)+(1+3 \alpha_3+\alpha_5)^2\Lambda]}}\right)\Lambda_3 t^2\nonumber \\
&&-\frac{1}{2(1+3 \alpha_3+\alpha_5)}\Lambda_3 r^2,\nonumber \\
A_0&=& -\frac{Q_0}{2}\Lambda_3 r^2.
\eea
In this case, the relationship (\ref{q0delta0}) between $\Delta_0$ and $Q_0$, gets modify to
\be\label{delta0}
\Delta_0=\frac{(1+3 \alpha_3+\alpha_5)^4 [(1+3 \alpha_3+2 \alpha_5)+(1+3 \alpha_3+\alpha_5)^2\Lambda]}{3
   Q_0^2 (2+3 \alpha_3+\alpha_5)^2 (1+3
   \alpha_3+\alpha_5)^4+(2+3 \alpha_3+\alpha_5)^4 [(1+3 \alpha_3+2 \alpha_5)+(1+3 \alpha_3+\alpha_5)^2\Lambda]},
\ee
and the AdS bound (\ref{adsbound}) to
\be
Q_0^2<\frac{(2 + 3 \alpha_3 + \alpha_5)^2\Big|(1+3 \alpha_3+2 \alpha_5)+(1+3 \alpha_3+\alpha_5)^2\Lambda\Big|}{3(1 + 3 \alpha_3 + \alpha_5)^4}.
\ee

\section{Decoupling limit Lagrangian with spherical symmetry}\label{AppC}
Let us consider the following Ansatz
\begin{equation}\label{ansatzall}
h_{\mu \nu}= f(t,r)\, \eta_{\mu \nu},\qquad
\pi =  \pi(t,r), \qquad
A_{\mu} = ( A_t(t,r), A_r(t,r), 0, 0).\,
\end{equation}
By inserting this into the action (\ref{genlag}), one can take the decoupling limit to obtain the following Lagrangian
\be\label{laga3a4}
{\cal L}_{dec}=\frac{r^2}{2}\left[ \frac{F_1( \pi )}{m^2}+{\cal L}_R^{(2)}({f})+F_2( \pi){f}(t,r)+F_3(\pi,A_t,A_r)\right]\,,
\ee
where we have set $\Lambda_3$ to one. The first term, $F_1$, is a total derivative and ${\cal L}_R^{(2)}({f})$ is the quadratic expansion of Einstein piece with a bare cosmological constant $\Lambda$, which reduces to
\be
{\cal L}_R^{(2)}({f})=\frac{3}{2r} \left(f'^2-\dot{f}^2\right)-4\Lambda f.
\ee
The term with $F_2$ is simply given by \cite{deRham:2010ik}
\bea
F_2&=&\sum_{i=1}^3 \Tr(X^{(i)})=-3\left(\ddot{\pi} - \pi'' - 2\frac{\pi'}{r}\right) +
   \left[(1 +3 \alpha_3)^2 - \alpha_5^2\right] \left[2(\dot{\pi}'^2 - \ddot{\pi} \pi'')-(\ddot{\pi}-\pi'')\frac{\pi'}{r}\right] \frac{\pi'}{r}\nonumber \\&& \hspace{2.5cm} + 4(1 + 3 \alpha_3) \left[\dot{\pi}'^2 + \frac{\pi'}{r} \left(2 \pi'' + \frac{\pi'}{r}\right) -
      \ddot{\pi} \left(\pi'' + 2 \frac{\pi'}{r}\right)\right].
\eea
Finally, the interaction term between the vector and the scalar, can be obtained using the techniques described in section \ref{Lagsection}, but solving for $E$, and not only $\tr{E}$, since $M^{3/2}$ is also needed when cubic or higher powers of ${\cal K}$ are included. It is an straightforward generalisation to the $\alpha_3=\alpha_4=0$ case, and the final results is
\bea
F_3&=& \frac{1}{2 +\ddot{\pi}- \pi''}\Bigg\{2 \dot{A_r} A_t' \bigg[1 +\ddot{\pi}- \pi'' + \frac{2}{r} (-1 + 3 \alpha_3 (1 +\ddot{\pi}- \pi'')) \pi' \nonumber \\&& \hspace{3.cm}
    - \left[3\alpha_3 - (1 + 3 \alpha_3 + 9 \alpha_3^2 - \alpha_5^2) (1 +\ddot{\pi}- \pi'')\right] \frac{\pi'^2}{r^2}\bigg]
      \nonumber \\ && \hspace{2.cm} +
     \left(\dot{A_r}^2 +  A_t'^2\right)\left[1 + \frac{\pi'}{r} \left(2 +6\alpha_3 + \frac{\pi'}{r}\left(1+ 6 \alpha_3 + 9 \alpha_3^2 -\alpha_5^2 \right)\right)\right] \nonumber  \\ && \hspace{2.cm}
     - 2 A_r' \dot{A_t} \left(2 +\ddot{\pi}- \pi''\right) \left[1 + \frac{\pi'}{r} \left(6\alpha_3 + \frac{\pi'}{r}\left(1+ 3 \alpha_3 + 9 \alpha_3^2 -\alpha_5^2 \right)\right)\right] \Bigg\} \nonumber  \\&& -
      \frac{4}{r} A_r \Bigg\{A_r' \left(-1 +
        3 \alpha_3 \left(\ddot{\pi} - \frac{\pi'}{r}\right) + \left(1 + 3 \alpha_3 + 9 \alpha_3^2 - \alpha_5^2\right)\frac{\ddot{\pi}\pi'}{r}\right) \nonumber \\&& \hspace{1.cm}
  - (\dot{A_r} +  A_t') \dot{\pi}' \left(3\alpha_3 + \frac{\pi'}{r}\left(1+ 3 \alpha_3 + 9 \alpha_3^2 -\alpha_5^2 \right)\right)
      \nonumber \\&& \hspace{1.cm} +
     \dot{A_t} \left(1 +\left(1 + 3 \alpha_3 + 9 \alpha_3^2 - \alpha_5^2\right) \frac{\pi'' \pi'}{r} + 3 \alpha_3 \left(\pi'' + \frac{\pi'}{r}\right)\right) \Bigg\}  \nonumber \\&&
    + \frac{2}{r^2} \bigg[1 + 3 \alpha_3 \left(-\ddot{\pi} + \pi''\right) +\left(1 + 3 \alpha_3 + 9 \alpha_3^2 - \alpha_5^2\right)\left(\dot{\pi}'^2 -\ddot{\pi}\pi''\right)\bigg] A_r^2.
\eea
One can check that the self-accelerating configuration (\ref{decsola3a4}) solves the equations of motion for this Lagrangian.

\subsection{Perturbations}\label{AppD}

In order to perform the perturbation analysis, we define the following perturbations
\be
h_{\mu \nu}= f(t,r)_0\, \eta_{\mu \nu} + \delta \chi_{\mu \nu}, \qquad
\pi = \pi_0 + \delta \pi(t,r), \quad
A_{\mu} = (A_{t0}(r) + \delta A_t(t,r), \delta A_r(t,r), 0, 0),
\ee
around the self-accelerating solution ($f_0,\pi_0,A_{t0}$) given by (\ref{decsola3a4}). One can decompose the tensor perturbations $\xi_{\mu\nu}$ in the same way as before (see Eq.~(\ref{xidec})), and show that do not contribute to the dynamics. The pure vector part is a total derivative as before, and only the vector-scalar mixing and pure scalar part play a role. The final action to second order in perturbations is
\be\label{S2_a3a4}
S^{(2)}=S^{(1)}_{VS}+S^{(1)}_{TS}+S^{S}_{VS}+S^{int}_{VS}+S^{S}_{TS},
\ee
where subscript represent the origin of such term (ST=scalar-tensor, SV=scalar-vector) and the upper label refers to pure scalar (S) or vector-scalar (int). The linear perturbations
\bea
S^{(1)}_{VS}\!\!&=&\!\!\frac{\alpha_5 \left[(1+3\alpha_3 +\alpha_5)^2-(2 +3 \alpha_3 + \alpha_5)^2 \sqrt{\Delta_0}\right]\left[1 +3 \alpha_3 + 2 \alpha_5 + (1 + 3 \alpha_3 + \alpha_5)^2 \Lambda\right] \delta\pi'r^3}{3 (1 + 3 \alpha_3 + \alpha_5)^3(2+3\alpha_3+\alpha_5)\sqrt{\Delta_0}}, \\ \nonumber
S^{(1)}_{TS}\!\!&=&\!\!\frac{\alpha_5 \left[(1 + 3 \alpha_3 + \alpha_5)^2 - (2 + 3 \alpha_3 + \alpha_5)^2 \sqrt{\Delta_0}\right] (r\delta\pi''
 + 2 \delta\pi') (1 + 3 \alpha_3 + 2 \alpha_5 + (1 + 3 \alpha_3 + \alpha_5)^2 \Lambda) r (r^2 -
    t^2)}{6 (1 + 3 \alpha_3 + \alpha_5)^3 (2 + 3 \alpha_3 + \alpha_5) \sqrt{\Delta_0}},
\eea
cancel after an integration by parts, as expected. The quadratic scalar parts reads
\bea
S^S_{VS}+S^S_{TS}=\int r^2 dt dr \frac{\alpha_5 \left(1+3 \alpha_3+2 \alpha_5+ (3
   \alpha_3+\alpha_5+1)^2\Lambda\right)}{\sqrt{\Delta_0} (3
   \alpha_3+\alpha_5+2)^2 (3 \alpha_3+\alpha_5+1)^2+(3
   \alpha_3+\alpha_5+1)^4}
   \bigg[\sqrt{\Delta_0} (3 \alpha_3+\alpha_5+2)^2 \delta\pi'^2\nonumber && \\
 \hspace{3cm} -(3 \alpha_3+\alpha_5+1)^2 \delta\dot{\pi}
   ^2\bigg]&&,
\eea
meanwhile the vector-scalar mixing term is
%
%
%
\be
S^{mix}_{VS}=-\frac{2 \alpha_5  \sqrt{((1 + 3 \alpha_3 + \alpha_5)^4 - (2 + 3 \alpha_3 + \alpha_5)^4 \Delta_0) (1 + 3 \alpha_3 +
    2 \alpha_5 + (1 + 3 \alpha_3 + \alpha_5)^2 \Lambda)}r^2}{\sqrt{3} (1 + 3 \alpha_3 +
   \alpha_5) ((1 + 3 \alpha_3 + \alpha_5)^2 + (2 + 3 \alpha_3 + \alpha_5)^2 \sqrt{\Delta_0}}
   \left(\delta A_t'\,  \delta\pi'
   -\delta \dot{A}_t \, \delta\dot{\pi}\right).
\ee
Therefore, the action in the main text (\ref{actionperts}) is obtained from the last two contributions, but expressed in terms of $Q_0$ instead of $\Delta_0$, which is achieved by equation (\ref{delta0}).

\end{appendix}

\end{document}